\documentclass[a4paper,11pt]{article}
\pdfoutput=1 

\usepackage{jcappub} 
\usepackage{overpic}
\usepackage[T1]{fontenc} 
\usepackage{color}
\title{\boldmath \duster: Dust Extinction with Photometric Red Galaxies}

\newcommand{\duster}{{\sc DustER}}
\newcommand{\cardinal}{{\sc Cardinal}}
\newcommand{\redmapper}{{\sc redMaPPer}}
\newcommand{\redmagic}{{\sc redMaGiC}}
\newcommand{\mdtrue}{m^{\rm (D)}_{\rm true}}
\newcommand{\mtrue}{m_{\rm true}}
\newcommand{\Igal}{I_{\text{gal}}}
\newcommand{\mlim}{m_{\text{lim}}}

\newcommand{\nside}{N_{\mathrm{side}}}
\newcommand{\avg}[1]{\langle #1 \rangle}
\newcommand{\srho}{\sigma_{\rm \rho}}
\newcommand{\drhoD}{\delta_\rho^{\rm D}}
\newcommand{\drhoC}{\delta_\rho^{\rm C}}
\newcommand{\calE}{{\mathcal{E}}}

\author[a,1]{I. Espinoza-Bustamante,\note{Corresponding author.}}
\author[b,c]{E. S. Rykoff,}
\author[a]{E. Rozo,}
\author[d,e,f]{C. To,}
\author[g]{and S. S. Boruah}

\affiliation[a]{Department of Physics, University of Arizona, Tucson, AZ 85721, USA}
\affiliation[b]{SLAC National Accelerator Laboratory, 2575 Sand Hill Rd., Menlo Park, CA 94025, USA}
\affiliation[c]{Kavli Institute for Particle Astrophysics and Cosmology, SLAC National Accelerator Laboratory, 2575 Sand Hill Rd., Menlo Park, CA
94025, USA}
\affiliation[d]{Astronomy and Astrophysics Department, The University of Chicago, Chicago, IL 60637, USA}
\affiliation[e]{Kavli Institute for Cosmological Physics, The University of Chicago, Chicago, IL 60637, USA}
\affiliation[f]{NSF-Simons AI Institute for the Sky (SkAI), 172 E. Chestnut St., Chicago, IL 60611, USA}
\affiliation[g]{Department of Physics and Astronomy, University of Pennsylvania, Philadelphia, PA 19104, USA}
\emailAdd{ivanespinoza@arizona.edu}
\emailAdd{erykoff@stanford.edu}
\emailAdd{erozo@arizona.edu}
\emailAdd{chto@uchicago.edu}
\emailAdd{supranta@sas.upenn.edu}

\abstract{We present \duster, a new method for reconstructing dust reddening maps from photometric galaxy survey data. \duster\ uses red galaxies as “standard crayons”. That is, it relies on the mean color of red-sequence galaxies being independent of their position in the sky. Given the galaxy's redshift, the difference between a galaxy’s observed color and its expected color measures the reddening by dust along the galaxy's line of sight. When using photometric galaxy survey data, the method must rely on photometric redshifts, which, combined with selection effects, bias the recovered dust maps. \duster\ uses a novel approach to characterize and correct for these biases by leveraging the known statistical properties of the distribution of dust around the Milky Way. Testing on simulations, we find that the recovered dust maps are contaminated by large-scale structure. However, when we use the \duster\ map to deredden and select red-sequence galaxies, we find the resulting two-point function is not contaminated by dust.
}

\begin{document}
\maketitle
\flushbottom

\section{Introduction}

Interstellar dust in the Milky Way is concentrated in a thin layer along the Galactic disk. Because we observe from within this layer, light from any extragalactic source must travel through part of it before reaching us. At optical wavelengths, dust grains both absorb and scatter light, making sources appear fainter and redder than they are. Consequently, the photometry of extragalactic sources needs to be corrected for the reddening and attenuation accumulated along the line of sight.  Performing these corrections requires two things: 1) the total dust column along each line of sight; and 2) the wavelength dependence of dust extinction, i.e., the reddening law.  This paper addresses the first of these. For this purpose only the projected dust column matters: whereas studies of Galactic structure require the three-dimensional distribution of dust, extragalactic photometry is affected equally by dust anywhere along the line of sight.

There are several methods for determining the distribution of dust in the Galaxy. The first is through its emission in the infrared \citep[e.g.][henceforth referred to simply as SFD]{Schlegel1998}. However, these can be contaminated by large-scale structure (LSS) due to confusion with the Cosmic Infrared Background (CIB) \cite{Chiang2019}. The addition of spatial templates can be used to decorrelate the dust from the CIB emission at the cost of added variance on the map \cite{Planck_colab_2016}. A second approach relies on tracers whose abundance correlates with dust, e.g. HI \cite{Lenz2017} and Polycyclic Aromatic Hydrocarbons (PAH) \cite{Meisner2013}. As with IR-based maps, PAH-based maps exhibit residual large-scale structure contamination from the diffuse PAH background, whereas HI maps appear to be free from large-scale structure contamination \cite{Chiang2019}. However, HI-based maps assume a fixed dust-to-gas ratio and miss dust associated with molecular gas, which restricts them to regions of low column density ($E(B-V) \lesssim 45$~mmag)~\cite{Lenz2017}. The third approach relies on measuring dust reddening directly, using sources of known intrinsic color: it measures the color excess of the sources themselves, which is the quantity needed to correct photometry, rather than inferring reddening from dust emission or from the abundance of a tracer through an assumed conversion. In this method the sources can be either stars \cite{Berry2012,Schlafly2014} or galaxies \citep{Peek2010}.  When using stars, if the distance to the calibration stars is known, then one can reconstruct three-dimensional maps of the Milky Way dust \cite{Green2015}, a method that is particularly powerful in combination with Gaia data \cite{Leike2019,Green2019,Guo2021}.  However, because of the disk structure of our Galaxy, the resolution of these maps decreases with increasing Galactic latitude, and the maps are limited to a distance of $\approx 5\ \text{kpc}$~\cite{Schlafly2014}. Further, and somewhat surprisingly, stellar-based dust maps also suffer from contamination by large-scale structure, likely due to quasar/stellar confusion \citep{Chiang2019}.

When the sources of known intrinsic color are galaxies, the method is referred to as the ``standard crayon'' method, first proposed by Peek \& Graves~\cite{Peek2010} and improved by Bravo et al.~\cite{Bravo2021}. Conceptually, in this method one identifies a population of extragalactic objects of known Spectral Energy Distribution (SED). By comparing the observed colors of these objects to their expected color, one can determine the amount of reddening due to dust along the line of sight to the object. Relative to IR or tracer-based methods, the standard crayon approach has the advantage of measuring dust reddening directly.  Relative to the stellar-based methods, standard crayon maps benefit from uniform resolution and the guarantee that they are sensitive to the entirety of the dust column along each line of sight, since the sources lie well beyond the Galaxy. However, there is an important disadvantage: the standard crayon method is limited in resolution by the availability of extragalactic standard crayon sources. The standard crayon method also has another apparent disadvantage, namely, it requires the adoption of a reddening law a priori. However, every dust map requires a reddening law before it can be used to correct photometry: emission- and tracer-based maps are calibrated to $E(B-V)$, and converting $E(B-V)$ into the extinction in any other band requires a reddening law. Consequently, this is not a significant drawback.

The main challenge of the standard crayon method is identifying the crayons. The original Peek \& Graves analysis relied on early-type galaxies that were selected based on their spectroscopic features. The SED for these galaxies was then calibrated using spectra from galaxies along lines of sight with negligible dust column density. The most important drawback of the Peek \& Graves approach is the reliance on spectroscopic data sets, which limits the resolution of the resulting dust maps, and prevents us from applying their algorithm on purely photometric data sets. 

Bravo et al. proposed a generalization of the Peek \& Graves methodology that simultaneously addresses both of the above drawbacks. They propose to bin galaxies in photometric redshift bins, assuming the latter are unbiased and have low scatter ($\sigma_z/(1+z)=0.02$). As in the Peek \& Graves analysis, the true mean color of galaxies in a given photometric redshift bin is determined from galaxies at locations with negligible dust column density. They then measure the color offset relative to the predicted galaxy color, and use this offset to estimate the dust column along the line of sight to each galaxy. Averaging over many galaxies yields an estimate of the dust column in each sky pixel. In practice, however, photometric redshifts are generically neither Gaussian nor unbiased, presenting a challenge to this method. \duster\  addresses this challenge by relying on \redmagic\ galaxies~\cite{Rozo2016} which are selected precisely because of the quality of their photometric redshifts. Unlike previous approaches, which calibrate the crayons on lines of sight with negligible dust, \duster\ calibrates the red sequence on the raw photometry of the full survey footprint, so that no dust-free calibration region is required. This choice biases the initial reconstruction, which is also noisy. \duster\ therefore includes a Wiener filtering step and a statistical debiasing step, which reduce the noise and correct for these biases. We conclude by testing the resulting maps for large-scale structure contamination, finding that a residual contamination is present and that a linear template-based cleaning does not remove it. Nevertheless, when we use the \duster\ map to deredden and select red-sequence galaxies in the  simulation, the clustering of the resulting sample agrees with that obtained using the true dust map well within full-sky cosmic variance. That is, in our simulations, dereddening with \duster\ does not measurably bias the clustering of red-sequence galaxies.

The layout of this paper is as follows. In Section~\ref{sec:conventions} we introduce notation and conventions. Section~\ref{sec:simulations} describes the \cardinal\ catalog and the procedure used to incorporate dust reddening and photometric noise. In Section~\ref{sec:algorithm} we present the \duster\ algorithm. Section~\ref{sec:simulationtests} contains validation tests using the dust-modified \cardinal\ catalog, including controlled tests of how dust and survey depth affect red-galaxy selection, and an analysis of large-scale structure imprints and mitigation strategies. We conclude in Section~\ref{sec:sumanddisc}.


\section{Dust Conventions}
\label{sec:conventions}

Because \duster\ infers the dust column from color excesses measured through broadband filters, it is important to be precise about what quantity a dust map represents. The Milky Way’s dust distribution is usually described using maps labeled as $E(B-V)$, where $E(B-V)$ is usually described as the excess Johnson $B-V$ color caused by dust reddening. In this work we distinguish between the observed color excess and the underlying dust density field, which we denote as the \textit{extinction field} $\calE$. Concretely, we take $\calE$ to be proportional to the dust column density along the line of sight, with the proportionality constant fixed by the optically thin relation $E(B-V) \equiv \calE$. Note that at high attenuation this linear relation breaks down for broadband filters due to the wavelength dependence of dust attenuation integrated over the bandpass (see equation~\ref{eq:Ab_second_order}). As a result, the color excess $E(B-V)$ is no longer proportional to the dust column density $\calE$, and the equality $E(B-V) = \calE$ is violated.

Of course, it is generally the case that one is interested in the dust density $\calE$ rather than the actual $B-V$ color excess, so the latter has become synonymous with dust density.  For example, the SFD maps are maps of $\calE$ scaled so that $E(B-V)=\calE$ in the optically thin regime~\cite{Schlegel1998}. Although SFD makes the distinction clear, this notation turns the physically correct statement $E(B-V)\neq\calE$ into the seemingly self-contradictory $E(B-V)\neq E(B-V)$ in regions of high column density. We therefore always denote the dust column by $\calE$.

To illustrate the difference between $E(B-V)$ and $\calE$, consider a source with flux $F(\lambda)$, and let $S_b(\lambda)$ be the transmission fraction given a bandpass $b$. We will assume the source is attenuated by dust with a reddening law $a\tau(\lambda)$, where $\tau(\lambda)$ gives the shape of the wavelength dependence of the extinction curve $A(\lambda)$,
\begin{equation}
    \tau(\lambda) \equiv \frac{A(\lambda)}{A(1\ \mu{\rm m})},
\end{equation}
with the $1~\mu{\rm m}$ reference wavelength chosen by convention. The constant $a$ determines the attenuation strength of the dust, and is proportional to the dust column density. Given this setup, the source's observed flux is attenuated by a factor
\begin{equation}
    e_b = \frac{ \int d\lambda\ \lambda^{-1} S_b(\lambda)F(\lambda)\exp[-a\tau(\lambda)] }{ \int d\lambda\ \lambda^{-1}S_b(\lambda)F(\lambda) },
    \label{eq:eb}
\end{equation}
and therefore its attenuated magnitude is larger than its unattenuated magnitude by an additive constant
\begin{equation}
    A_b = - \frac{2.5}{\ln 10}\ln e_b. \label{eq:attenuation}
\end{equation}
Given equation~\ref{eq:attenuation}, we can calculate $A_b$, and hence $E(B-V)$, for any source, dust attenuation strength $a$, and reddening law $\tau(\lambda)$. In the limit of low attenuation, $\exp[-a\tau(\lambda)]\approx 1-a\tau(\lambda)$, which leads to
\begin{equation}
    A_b = \frac{2.5}{\ln 10}\,a\,\avg{\tau}_b,
\end{equation}
where
\begin{equation}
    \avg{\tau}_b = \frac{\int d\lambda\ \lambda^{-1} S_b(\lambda)F(\lambda)\tau(\lambda)}{\int d\lambda\ \lambda^{-1} S_b(\lambda)F(\lambda) }.
    \label{eq:tau_b}
\end{equation}
Adopting $F(\lambda)=\text{constant}$ in equation~\ref{eq:eb} and an \citet{o94} reddening law, we find $\avg{\tau}_B=2.8347$ and $\avg{\tau}_V=2.1654$, and therefore
\begin{equation}
    E(B-V) = 0.7267a.
\end{equation}
We \textit{define} the extinction field
\begin{equation}
    \calE = 0.7267a,
\end{equation}
so that $E(B-V)=\calE$ in the limit of low extinction. It is this quantity $\calE$ that is mapped out in traditional dust maps. Given $\calE$, we can then readily compute $E(B-V)$ via equation~\ref{eq:attenuation}.

The origin of the non-linearity between $E(B-V)$ and $\calE$ becomes clear when equation~\ref{eq:attenuation} is expanded to second order in $a$:
\begin{equation}
    A_b = \frac{2.5}{\ln 10}\left[a\avg{\tau}_b
          - \frac{a^2}{2}\,\sigma^2_{\tau,b} + \mathcal{O}(a^3)\right],
    \qquad
    \sigma^2_{\tau,b} \equiv \avg{\tau^2}_b - \avg{\tau}_b^2,
    \label{eq:Ab_second_order}
\end{equation}
where $\avg{\tau^2}_b$ is defined as in equation~\ref{eq:tau_b} with $\tau$ replaced by $\tau^2$. The leading-order correction is set by the variance of the extinction curve across the band, computed with the same weighting as equation~\ref{eq:tau_b}, and all corrections vanish if $\tau(\lambda)$ is constant across the band. At fixed central wavelength, this variance grows with the width of the bandpass: approximating $\tau(\lambda)$ as linear across the band gives $\sigma^2_{\tau,b}\approx[\tau'(\lambda_b)]^2\sigma^2_{\lambda,b}$, where $\lambda_b$ and $\sigma^2_{\lambda,b}$ are the mean and the variance in wavelength across the band. Narrow bands therefore remain linear in $\calE$ up to high attenuation, whereas very broad bands are strongly non-linear. 

\begin{figure}[tbp]
\centering 
\includegraphics[width=.65\textwidth]{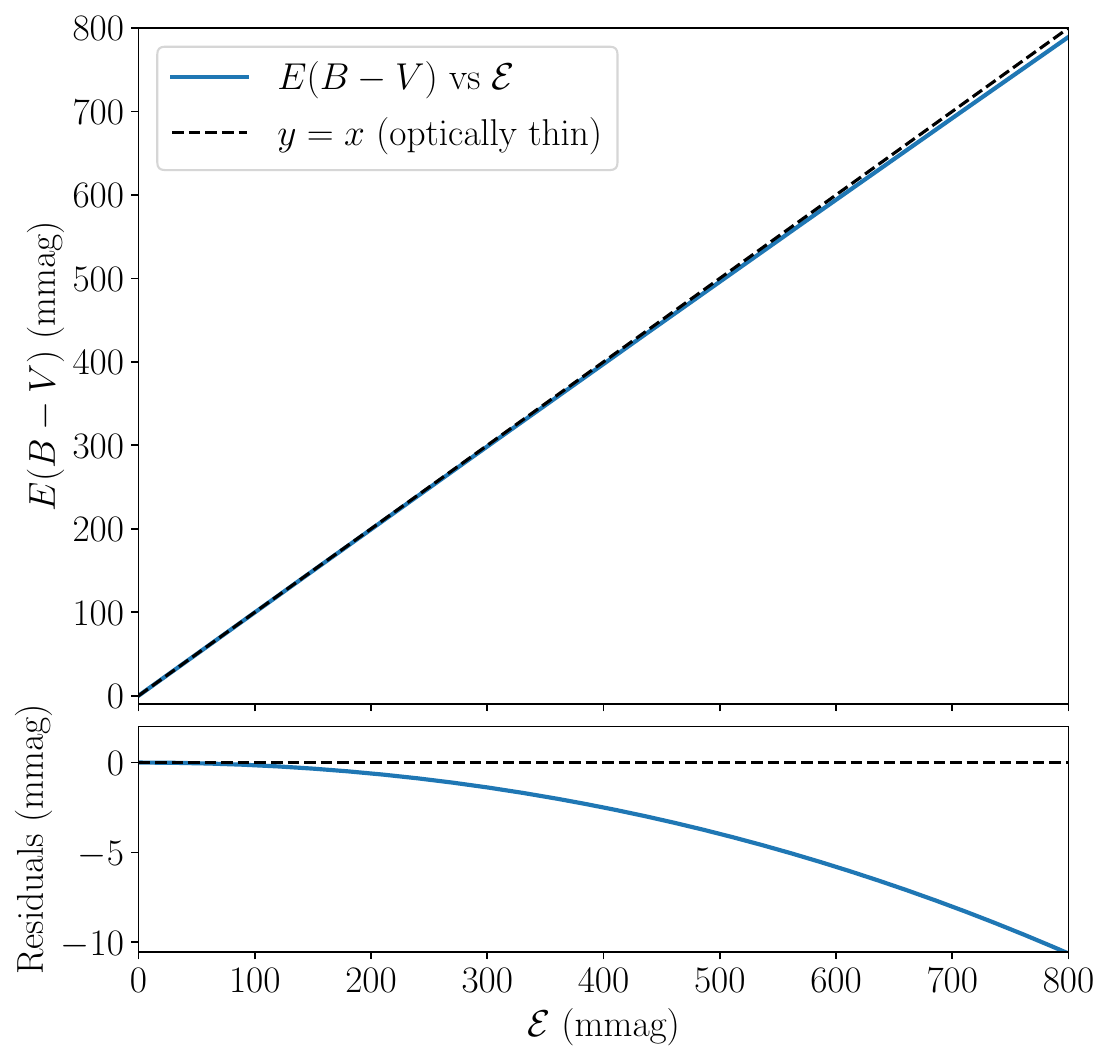}
\caption{Color excess $E(B-V)$ as a function of the extinction field $\mathcal{E}$, computed for the Johnson $B$ and $V$ bandpasses with $F(\lambda)=\text{constant}$ in equation~\ref{eq:eb} and an O’Donnell reddening law \cite{o94}. The dashed line corresponds to the optically thin limit in which $E(B-V)=\mathcal{E}$. 
At higher extinction, the linear relation breaks down due to the wavelength-dependent 
dust attenuation integrated over the bandpass, highlighting the importance of 
distinguishing these two quantities.}
\label{fig:calEvsEBV}
\end{figure}

Figure~\ref{fig:calEvsEBV} shows the color excess $E(B-V)$ as a function of the extinction field $\calE$. Also shown for reference is the $y=x$ line. We see the two curves match at low extinction, but differ at high extinction. This is why it is necessary to differentiate between $E(B-V)$ and $\calE$, since $E(B-V)$ is not always linear in $\calE$. In this work, we are concerned with mapping out the extinction field $\calE$.

\section{Simulations}
\label{sec:simulations}

\duster\ is an algorithm for recovering the distribution of Milky-Way dust using the colors of photometrically-selected red galaxies. We test \duster\ using \cardinal, a simulated photometric data set designed to match DES Y3 data \citep{cardinal}.  This synthetic data set is useful not only for testing \duster, but also for illustrating each of the steps in the algorithm as it is described. Readers unconcerned with how the simulation data is generated need to know only two things: 1) our simulated data set takes the form of a photometric galaxy catalog in which every galaxy has been attenuated by dust using either the SFD or \textit{Planck} GNILC map (see Section~\ref{sec:adding_dust}), and then ``observed'' in a way that mimics the DES Y3 data; and 2) the galaxies in the galaxy catalog are clustered in a way that mimics DES Y3 data.  If the reader is willing to believe these two properties of our synthetic data set, they can skip to Section~\ref{sec:algorithm} for a description of the \duster\ algorithm.  We use simulation tests to assess the robustness of our results to systematic uncertainties.

\subsection{The Cardinal Simulation}

We test the \duster\ algorithm using synthetic data sets generated from the \cardinal\ simulation \cite{cardinal}. \cardinal\ populates a $10{,}000\ \deg^2$ lightcone simulation with galaxies of $r$-band magnitude $r<27$ over the redshift range $z\in [0.0,2.35]$.  The lightcone is constructed using three N-body simulations with sizes $1.05^3$, $2.60^3$, and $4.00^3\ h^{-3}\ \mathrm{Gpc}^3$, and mass resolution of $3.3\times10^{10}$, $1.6\times10^{11}$, and $5.9\times10^{11}h^{-1}M_\odot$. The N-body simulations are generated using a $\Lambda$CDM cosmology with $\Omega_m=0.286$, $\Omega_b=0.047$, $\sigma_8 = 0.82$, $n_s=0.96$, $h=0.7$, $A_s=2.145\times10^{-9}$, and three massless neutrino species with $N_\mathrm{eff}=3.046$. Galaxies are assigned to dark-matter particles by giving them galaxy positions, velocities, and absolute magnitudes based on their Lagrangian overdensities. The relation between galaxy properties and the overdensities is derived from a subhalo abundance matching model (SHAM) constrained by the galaxy-galaxy correlations and group-galaxy correlations measured in SDSS. The resulting model approximately reproduces the luminosity functions, star formation histories, and observed luminosity- and color-dependent clustering of galaxies. Additionally, \cardinal\ includes orphan subhalos as constrained by the galaxy-galaxy and group-galaxy correlations.  \cardinal\ then uses a conditional abundance matching algorithm to assign every galaxy an SED from a Sloan Digital Sky Survey (SDSS) galaxy, which then is used to calculate the galaxy's $ugriz$ magnitudes. For a more detailed description of \cardinal\ we refer the reader to \cite{cardinal}.

\subsection{Adding Dust}
\label{sec:adding_dust}

The magnitudes of the galaxies in the \cardinal\ galaxy catalog are dust-free.  As light from distant sources propagates through the dust clouds of the Milky Way, some fraction of the light will be absorbed or scattered.  We refer to the magnitude of a galaxy observed through dust at infinite signal-to-noise as the true \it dusty \rm magnitude of the galaxy, $m_{{\text{true}}}^{\text{(D)}}$.  This is related to the true magnitude of the galaxy via
\begin{equation}
    \mdtrue = \mtrue+a_m \calE,
    \label{eq:reddeningmags}
\end{equation}
where $\calE$ is the extinction field, and $a_m$ is a band-dependent extinction coefficient.\footnote{In practice, the extinction coefficient is wavelength dependent, and therefore will vary somewhat with galaxy spectral type. The effect is small, and we will ignore it for the rest of this work.} Note that because we assume linearity, our analysis must be limited to low-extinction regions of the sky.  For the purposes of our fiducial simulation, we will assume the extinction field is given by the famous Schlegel, Finkbeiner, and Davis dust map \cite{Schlegel1998}, referred to simply as the SFD map. To test whether our results depend on this choice, we also generate a second catalog in which the extinction field is given by the Planck GNILC dust map \citep{Planck_colab_2016} (henceforth P16-GNILC); the two catalogs are otherwise identical. Unless stated otherwise, all results refer to the fiducial SFD-reddened catalog. The P16-GNILC catalog is used only in Section~\ref{sec:statperf} to test the robustness of \duster\ to the
input dust map.

The true dusty magnitudes are then used to generate observed magnitudes for each of the galaxies in \cardinal\ using the prescription described in \cite{Derose2019,Wechsler2022}.  Briefly, let $t_{\text{eff}}$ be the effective exposure time 
and $\Igal$ be the number of photons that impact the telescope in this time.  The expectation value of the number of photons is given by
\begin{equation}
    \langle I_{\text{gal}}\rangle=10^{-0.4(m_{{\text{true}}}^{\text{(D)}}-m_{\text{ZP}})}\times t_{\text{eff}},
\end{equation}
where $m_{\text{ZP}} = 30.0$ is the zero-point magnitude. We assume the survey depth in each band is characterized in terms of the $10\sigma$ limiting magnitude $\mlim$.\footnote{Our depth maps are constructed from observed (i.e., not dereddened) photometry; see Appendix~\ref{app:depth} for the impact of using dereddened depth maps.} From our Poisson assumption, it follows that the expectation value for the number of noise photons associated with a galaxy of magnitude $\mlim$ is
\begin{equation}
    \langle I_{\text{sky}}\rangle=\left(\frac{10^{-0.8(m_{\text{lim}}-m_{\text{ZP}})}\times t_{\text{eff}}}{100}-10^{-0.4(m_{\text{lim}}-m_{\text{ZP}})}\right)\times t_{\text{eff}}.
\end{equation}
The observed flux $F_{\text{obs}}$ of a galaxy can therefore be written as
\begin{equation*}
    F_{\text{obs}}=\frac{\text{Poisson}(\langle I_{\text{gal}}\rangle+\langle I_{\text{sky}}\rangle)-\langle I_{\text{sky}}\rangle}{t_{\text{eff}}},
\end{equation*}
where $\text{Poisson}(\lambda)$ denotes a random draw from a Poisson distribution with mean $\lambda$. Hence, the observed magnitudes and magnitude errors in the presence of dust are given by
\begin{equation*}
    m_{\text{obs}}=m_{\text{ZP}}-2.5\log_{10}(F_{\text{obs}}),
\end{equation*}
and
\begin{equation*}
    \sigma_{m,\text{obs}}=\frac{2.5}{\ln 10}\frac{\sigma_{F_\text{obs}}}{F_{\text{obs}}},
\end{equation*}
where $\sigma_{F_\text{obs}} = \sqrt{\langle I_{\text{gal}}\rangle+\langle I_{\text{sky}}\rangle}/t_{\text{eff}}$. Unfortunately, the background subtraction can lead to negative fluxes, and therefore ill-defined magnitudes.  Consequently, we use arcsinh magnitudes (sometimes referred to as luptitudes) as proposed in \cite{Lupton1999}. With this convention, one has
\begin{equation*}
    \mu = -\frac{2.5}{\ln 10}\left[\text{arcsinh}\left(\frac{F_\text{obs}}{2bF_0}\right)+\ln b \right],
\end{equation*}
and
\begin{equation*}
    \sigma_{\mu}=\frac{2.5}{\ln10}\frac{\sigma_{F_\text{obs}}}{\sqrt{F_\text{obs}^2+(2bF_0)^2}},
\end{equation*}
where $F_0$ is the flux associated with a galaxy with $m=0$ and $b$ is defined as a \textit{softening parameter} which determines the flux level at which linear behavior sets in. In Table~\ref{tab:exctinctioncoeffandsoftparams} we provide the numerical values of the extinction coefficients and softening parameters used in this work. Finally, we generate synthetic luptitudes and luptitude errors for every \cardinal\ galaxy as described above.  The resulting galaxy catalog is the input to \duster.
\begin{table}[tbp]
\centering
\begin{tabular}{|c|c|c|}
\hline
\textbf{Band}&$a_m$&$b$\\
\hline
$g$&$3.186$&$3.27\times 10^{-12}$\\
\hline
$r$&$2.140$&$4.83\times 10^{-12}$\\
\hline
$i$&$1.569$&$6.0\times 10^{-12}$\\
\hline
$z$&$1.196$&$9.0\times 10^{-12}$\\
\hline
\end{tabular}
\caption{Extinction coefficients and softening parameters with their respective photometric bands used in this work. The extinction coefficients are the fiducial DES DR1 values~\cite{DESDR1}.  These extinction coefficients are appropriate for when dust density is measured in its standard units of $\mathcal{E}$.}
\label{tab:exctinctioncoeffandsoftparams}
\end{table}

\section{The \duster\ Algorithm}
\label{sec:algorithm}

The fundamental insight/assumption of the \duster\ algorithm is that, at fixed redshift, all red-sequence galaxies follow a tight color--magnitude relation with small intrinsic scatter.  Since dust extinction is stronger in bluer bands, dust extinction reddens the colors of galaxies.  If one is able to calibrate the average color of red-sequence galaxies as a function of redshift, then one can use coherent fluctuations in the color of these galaxies to map out fluctuations in the dust density of the Milky Way.  Consequently, the first step towards mapping dust fluctuations is selecting samples of red-sequence galaxies of known redshift and colors.

\subsection{Calibration of the Color--Redshift Relation}

We calibrate the color--redshift relation of red-sequence galaxies using the \redmapper\ cluster finding algorithm \cite{Rykoff2014}. \redmapper\ runs in two stages. The first is a calibration stage in which the color--redshift relation of red galaxies is empirically self-calibrated.  The second consists of a cluster finding stage in which the calibrated model is used to identify galaxy clusters and measure their richness.  Our results rely exclusively on the output of the first stage, which we briefly describe below.

\redmapper\ begins by identifying bright spectroscopic galaxies, which we refer to as ``seeds''. It then identifies the seed galaxies that reside within overdensities of similarly colored galaxies.  The color of the seed galaxy is then used in \redmapper\ to identify likely cluster members around the seed.  Finally, \redmapper\ simultaneously fits all high-likelihood cluster members around seeds to derive the redshift-dependent color model of red-sequence galaxies as a function of redshift. This model describes the red sequence at each redshift through its intercept, its slope (the tilt of the red sequence in the color--magnitude diagram), and its intrinsic scatter~\cite{Rykoff2014}. With this calibrated color model in hand, the procedure is iterated three times, resulting in the final color--redshift relation used to select red galaxies as described in Section~\ref{sec:redgals}. 

We emphasize that we calibrate our red-sequence model using raw (i.e. not dereddened) galaxy magnitudes.  This is done to ensure the \duster\ dust maps are not informed by the reddening values from existing dust maps.  While this choice implies our recovered dust maps are biased at first, we demonstrate below that we can remove this bias as part of our algorithm.

\subsection{Red Galaxy Selection}
\label{sec:redgals}

Our algorithm to select red galaxies is a simplified version of the \redmagic\ algorithm \cite{Rozo2016}. We first fit \it every \rm galaxy with the \redmapper\ red-sequence model to determine a photometric redshift estimate $z_{\rm red}$ for each galaxy. This is done by minimizing 
\begin{equation}
    \chi^2(m,z)=(\mathbf{c}-\langle\mathbf{c}|m,z\rangle)^T(\mathbf{C}_{\mathrm{int}}(z)+\mathbf{C}_{\mathrm{obs}})^{-1}(\mathbf{c}-\langle\mathbf{c}|m,z\rangle),
\end{equation}
where $\mathbf{c}$ is the color vector of a galaxy, $\langle\mathbf{c}|m,z\rangle$ is the \redmapper\ color model, $\mathbf{C} _{\mathrm{int}}$ is the covariance characterizing the intrinsic width of the red sequence, and $\mathbf{C} _{\mathrm{obs}}$ is the covariance matrix describing the photometric errors in the galaxy colors. After assigning each galaxy a photometric redshift $z_\mathrm{red}$, we select galaxies with $\chi^2(m,z_\mathrm{red})\leq8$ and $\ell(m,z_\mathrm{red})\geq0.5$, where 
\begin{equation}
    \ell(m,z) \equiv \frac{L}{L_\star} = 10^{-\frac{2}{5}[m-m_*(z)]}.
\end{equation}
The value $m_*(z)$ is set to $i = 17.85$ at $z = 0.2$ to match \redmapper. The evolution of $m_*(z)$ is computed using the Bruzual \& Charlot \cite{Bruzual2003} stellar population synthesis code as implemented in the EzGal Python package.\footnote{http://www.baryons.org/ezgal} Note that as with the red-sequence calibration, our selection of red galaxies is performed using \it raw \rm (i.e. not dereddened) photometry.

\subsection{The Raw \duster\ Map}

We now use the spatial modulation of color residuals of red-sequence galaxies to construct what we refer to as the ``raw'' \duster\ map.  We can choose to arbitrarily normalize the dust density so that it has a mean of one. It follows then that $\rho = \mathcal{E}/\overline{\mathcal{E}}$ where $\overline{\mathcal{E}}$ denotes the average across all pixels in the survey. This normalization is useful for decoupling the physical dust density from the dust law used to relate dust column density to dust attenuation.

With this reparameterization in hand, the change in a galaxy magnitude is given by
\begin{equation}
    \Delta m = a_m\overline{\mathcal{E}}\rho,
\end{equation}
where $a_m$ is the band-dependent extinction coefficient introduced in Section~\ref{sec:adding_dust}. Consequently, the change in the galaxy color $m-n$ due to dust is
\begin{equation}
    \Delta c_{mn}=\Delta m-\Delta n = (a_m-a_n)\overline{\mathcal{E}}\rho.
\end{equation}
Using our selection of red galaxies, we model the dust density over a sky pixel as constant, and we measure the color $c_{mn,\text{obs}}$ of every galaxy in the pixel. To ensure that the dust estimate per pixel is statistically reliable, we require at least 50 red-sequence galaxies within each pixel. This constraint sets the angular resolution of the map, corresponding to a HEALPix resolution of $N_{\mathrm{side}} = 128$ (pixel area $\approx 0.21\,\mathrm{deg}^2$), which balances spatial resolution against per-pixel noise. Pixels that do not meet this threshold are excluded from the analysis. We model the observed color of red-sequence galaxies in sky pixels as having Gaussian random noise of variance $\sigma^2_{mn}$ where $\sigma_{mn}$ is the sum in quadrature of the observed error in the color and the intrinsic spread of the red-sequence. The color expectation value is
\begin{equation}
    \langle c_{mn,\text{obs}}\rangle = c_{mn,\text{pred}}+(a_m-a_n)\overline{\mathcal{E}}\rho.
\end{equation}
To compute the predicted color $c_{mn,\text{pred}}$, we evaluate the predicted color for a red-sequence galaxy at the photometric redshift of the galaxy, as calibrated using \redmapper . Note that since the red-sequence calibration and photometric redshifts are themselves impacted by reddening,
the recovered dust maps are necessarily biased.

With these assumptions, the log-likelihood of our model is
\begin{align}
        \ln\mathcal{L}=&\,-\frac{1}{2}\sum_{\substack{\text{(galaxies in pixel)} \\ \text{and (all colors)}}}\frac{[c_{mn,\text{obs}}-c_{mn,\text{pred}}-(a_m-a_n)\overline{\mathcal{E}}\rho]^2}{\sigma^2_{mn}
        },  
\end{align}
where the sum runs over colors $g-r$, $r-i$, and $i-z$. For simplicity, we neglect the covariance between colors induced by shared photometric bands. Defining
\begin{equation}
    \overline{\rho}_{mn}=(c_{mn,\text{obs}}-c_{mn,\text{pred}})/[(a_m-a_n)\overline{\mathcal{E}}],
\end{equation}
and
\begin{equation}
    \tilde{\sigma}_{mn}^2=\sigma_{mn}^2/[(a_m-a_n)\overline{\mathcal{E}}]^2,
\end{equation}
we find that the maximum likelihood point is
\begin{equation}
    \rho_{\text{raw}}=\frac{\sum w_{mn}\overline{\rho}_{mn}}{\sum w_{mn}},
\end{equation}
and $w_{mn}=1/\tilde{\sigma}_{mn}^2$. We refer to $\rho_{\text{raw}}$ as the raw \duster\ map.  The left panel in Figure~\ref{fig:i} shows the dust density from the raw \duster\ map against the true dust density in the simulation.  As expected, the raw dust map is both noisy and biased, failings which we will address in the next two sections.


\subsection{Wiener Filtering}

Before we address the bias in the raw \duster\ map, we will address another undesirable property of the maps: they are very noisy. Fortunately, the noise is not likely to be correlated from pixel to pixel: in the ideal case, it would be driven purely by the random realizations of the color draws in the galaxies from different pixels. For this reason, we expect the power-spectrum of the raw \duster\ map to exhibit white noise, which we can
suppress using standard de-noising techniques.

\begin{figure}[tbp]
\centering 
\includegraphics[width=.48\textwidth]{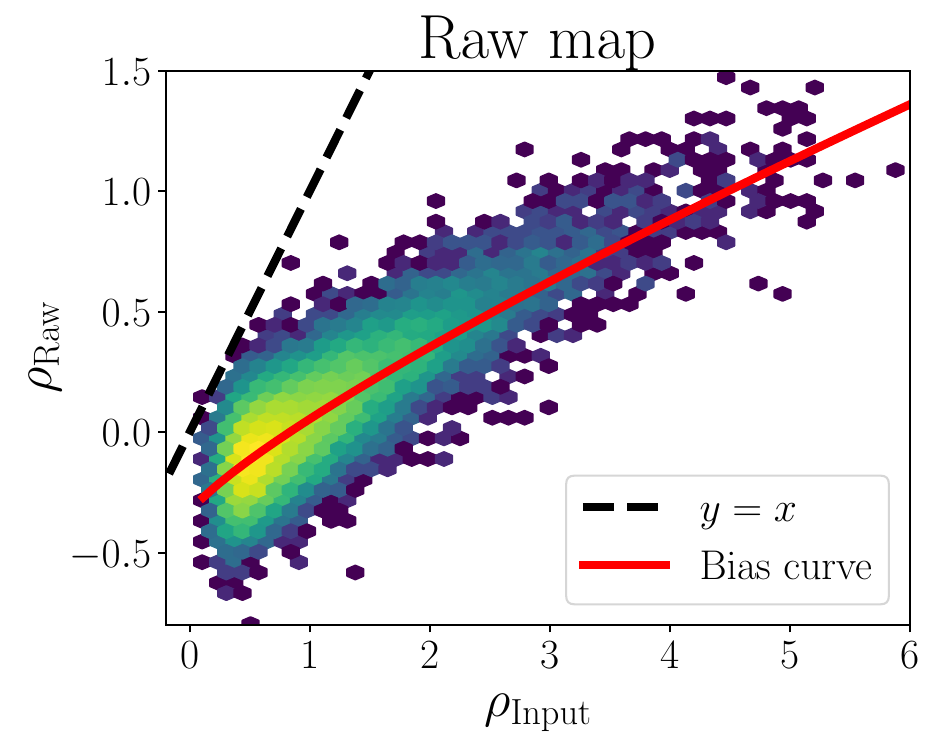}
\hfill
\includegraphics[width=.48\textwidth]{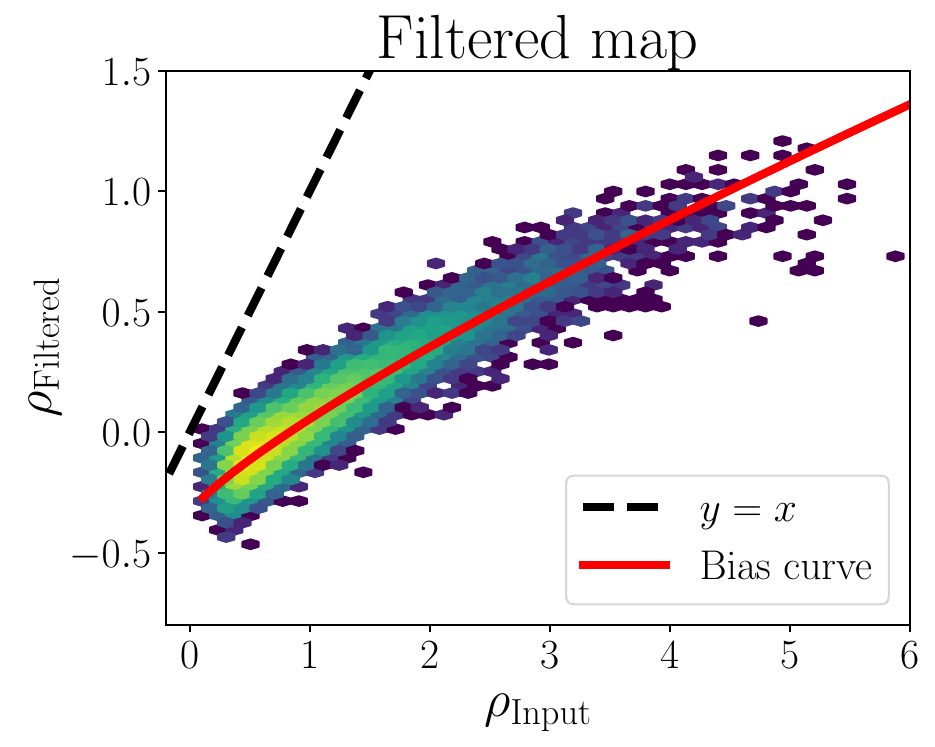}
\caption{\label{fig:i} {\bf Left: } Raw \duster\ map as a function of the true dust density in simulations (in our case, SFD). The raw \duster\ map is both noisy and biased. Because the red-sequence calibration is performed on raw photometry over the full survey footprint, which is not free of dust, the mean reddening of the footprint is absorbed into the color model. As a result, $\rho_{\rm Raw}$ is centered near zero and takes both positive and negative values. This offset is removed in the debiasing step (Section~\ref{sec:debiasing}), where the dust density is required to be non-negative through the one-point prior of Figure~\ref{fig:one_point_function}. The red curve is a fit to the expectation value of the raw map as a function of the input dust density as per equation~\ref{eq:avgbias}. The black dashed line corresponds to the $y=x$ line. {\bf Right:} Filtered \duster\ map as a function of the true dust density in simulations. Wiener filtering of the raw map reduces the noise of the recovered map, but the large systematic biases in the dust density remain.}
\end{figure}

The power spectrum $C_\ell^{\text{obs}}$ of the raw dust map is the sum of a signal contribution $C_\ell$ and a constant white noise term $C_n$. The Wiener filtered map $\tilde{\rho}$ is defined in terms of its spherical harmonic coefficients via
\begin{equation}
    \tilde{\rho}_{\ell m}=\rho_{\ell m}\frac{C_\ell}{C_\ell+C_n},
\end{equation}
where $\rho$ is the raw \duster\ map. To apply the Wiener filter we must estimate $C_\ell$ and $C_n$. After inspecting the power spectrum of the raw map, we model the power spectrum as the sum of a power-law and a constant noise term $C_n$, so that
\begin{equation}
    C_\ell^{\text{obs}}=k\left(\frac{\ell}{50}\right)^{-\eta}+C_n,
    \label{eq:cl_model}
\end{equation}
where $k$, $\eta$, and $C_n$ are model parameters. To find these
parameters, we compute $C_\ell^{\rm obs}$ from the mean-subtracted raw
map and fit equation~\ref{eq:cl_model} to individual multipoles with
$10\leq\ell\leq 3N_{\rm side}-1$ by unweighted least squares in
logarithmic space, i.e., by minimizing the cost function
\begin{equation*}
    C \equiv \sum_\ell\left[\ln C_\ell^{\rm obs}
    -\ln\left(k\left(\frac{\ell}{50}\right)^{-\eta}+C_n\right)\right]^2.
\end{equation*}
Fitting in logarithmic space prevents the fit from being dominated by the
low-$\ell$ multipoles, where the power is largest. We use the best-fit
values of $k$, $\eta$, and $C_n$ directly, ignoring their statistical
uncertainties. With our fits in hand, we apply the Wiener filter in
harmonic space, and then transform it back to real space to arrive at our
de-noised dust map. We will refer to the map $\tilde{\rho}$ as the
\it filtered \rm map. The right panel in Figure~\ref{fig:i} shows the filtered \duster\ map in our simulation as a function of the input dust density.  We can see the filtered map is significantly less noisy than the raw dust map (left panel), but the bias remains.

\subsection{Map Debiasing}
\label{sec:debiasing}

We model the bias in the filtered map as a power-law plus a constant,
\begin{equation}
    \langle \tilde{\rho}|\rho\rangle = \alpha \rho^\beta+\gamma,
    \label{eq:avgbias}
\end{equation}
where the parameters $\alpha$, $\beta$, and $\gamma$ characterize the systematic bias offset between the filtered map and the input map. We will further assume that the probability distribution $P(\tilde{\rho}|\rho)$ is Gaussian, with an unknown variance $\sigma^2$:
\begin{equation*}
    P(\tilde{\rho}|\rho)=\frac{1}{\sqrt{2\pi\sigma^2}}\exp\left(-\frac{1}{2}\frac{(\tilde{\rho}-\langle \tilde{\rho}|\rho\rangle)^2}{\sigma^2}\right).
\end{equation*}
The scatter $\sigma^2$ is itself treated as a model parameter.

\begin{figure}[tbp]
\centering 
\includegraphics[width=.65\textwidth]{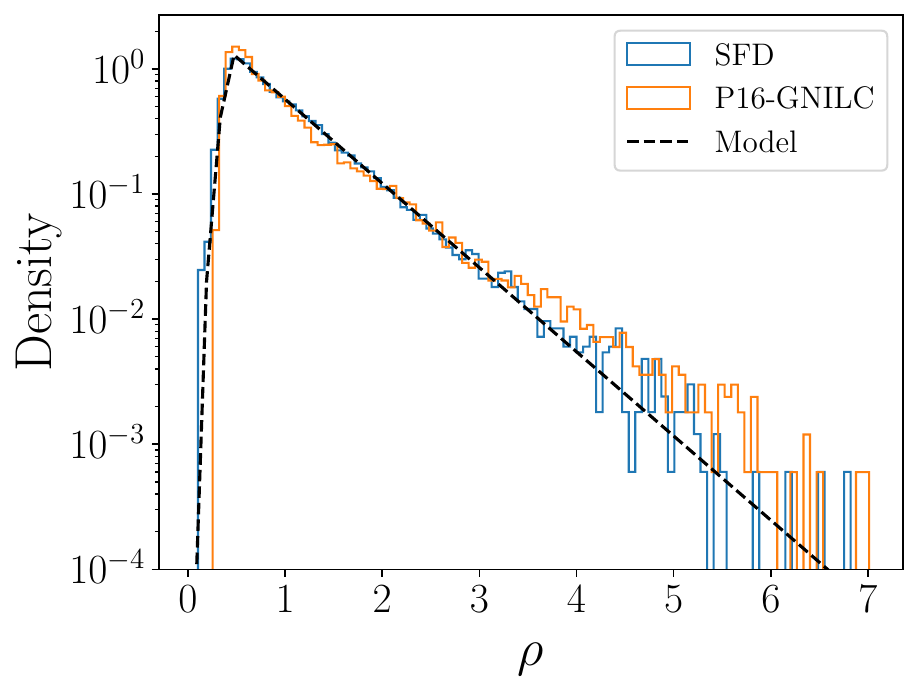}
\caption{A comparison of the one-point function of the SFD \citep[][blue]{Schlegel1998} and P16-GNILC \citep[][orange]{Planck_colab_2016} dust maps, and the best-fit model to SFD (dashed), adopted as the prior in \duster.}
\label{fig:one_point_function}
\end{figure}

Our challenge is to recover the parameters $\alpha$, $\beta$, $\gamma$, and $\sigma^2$ that characterize the true dust map $\rho$ and the filtered map $\tilde{\rho}$ without utilizing pixel-level information from any external dust map for calibration. To do so, we rely on the statistical properties of the dust maps.  Specifically, we characterize the one-point function of a dust map using the SFD and P16-GNILC dust maps. We find that after rescaling these maps to enforce a unit mean, the resulting one-point distributions are nearly identical (Figure~\ref{fig:one_point_function}), and can be accurately described via 
\begin{equation}
    P_{\text{dust}}(\rho|\rho_0,a,\rho_{\text{min}})=
    \begin{cases}
			B(\rho/\rho_0)^a & \text{if  $\rho\leq\rho_{\text{min}}$},\\
            B(\rho_{\text{min}}/\rho_0)^a \exp(-(\rho-\rho_{\text{min}})/\rho_0), & \text{otherwise},
    \end{cases}
\end{equation}
where
\begin{equation}
    B=\frac{a+1}{\rho_0(\rho_\text{min}/\rho_0)^a[a+1+(\rho_\text{min}/\rho_0)]},
\end{equation}
and $\rho_0$, $a$, and $\rho_\text{min}$ are unknown parameters. These parameters are constrained by the requirement that $\langle\rho\rangle=1$, which allows us to relate $a$ to $\rho_0$ and $\rho_\text{min}$. We find the final result is best parameterized in terms of $a$ and $u=\rho_\text{min}/\rho_0$, with
\begin{equation}
    \rho_0=\frac{a+2}{a+1}\frac{u+a+1}{u^2+(a+2)(u+1)}.
\end{equation}
The equations above define $\rho_0$ and $\rho_\text{min}$ in terms of $a$ and $u$, allowing us to fully specify the probability distribution $P_\text{dust}$ in terms of these two variables. We model the values of $\rho$ in the SFD and P16-GNILC dust maps as independent random draws from the above distribution, and recover nearly identical values for the parameters $a$ and $u$. Throughout this work, we adopt the best-fit parameters from SFD. \duster\ assumes the one-point function $P_\text{dust}$ is known, and we ignore the uncertainty in the recovered parameters $a$ and $u$.  Note that while the statistical properties of the SFD dust map do inform the statistical properties of the \duster\ map, the dust density recovered by \duster\ at any given location in the sky is independent of the dust density reported by SFD or P16-GNILC at that same location.


With a known probability distribution $P_\text{dust}$ in hand, we are now in a position to predict the probability distribution function of the filtered map $\tilde{\rho}$. Specifically, we have that
\begin{equation}
    P(\tilde{\rho}|\alpha,\beta,\gamma,\sigma^2)=\int\text{d}\rho\ P(\tilde{\rho}|\rho,\alpha,\beta,\gamma,\sigma^2)P_\text{dust}(\rho).
\end{equation}
The corresponding likelihood is
\begin{equation}
    \mathcal{L}(\alpha,\beta,\gamma,\sigma^2)=\prod_{\text{pixels}}P(\tilde{\rho}|\alpha,\beta,\gamma,\sigma^2).
\end{equation}
We sample this likelihood function to arrive at the posterior distribution for $\alpha$, $\beta$, $\gamma$, and $\sigma^2$. We then compute the mean and variance of the filtered map $\tilde{\rho}$ given $\rho$, marginalizing over the parameters $\alpha$, $\beta$, $\gamma$, and $\sigma^2$.  The first and second order moment of $\tilde\rho$ are given by 
\begin{equation}
    \langle\tilde{\rho}|\rho\rangle=\frac{1}{N}\sum_{i=1}^N(\alpha_i\rho^{\beta_i}+\gamma_i),
    \label{eq:mean}
\end{equation}
and
\begin{equation}
    \langle\tilde{\rho}^2|\rho\rangle=\frac{1}{N}\sum_{i=1}^N[(\alpha_i\rho^{\beta_i}+\gamma_i)^2+\sigma_i^2],
    \label{eq:mom2}
\end{equation}
where $(\alpha_i,\beta_i,\gamma_i,\sigma_i^2)$ is the $i^{\text{th}}$ point in the chain $\{\alpha,\beta,\gamma,\sigma^2\}$, and $N$ is the number of points in the chain. The variance is $\text{Var}(\tilde{\rho}|\rho)=\langle\tilde{\rho}^2|\rho\rangle-\langle\tilde{\rho}|\rho\rangle^2$. With the assumption of Gaussianity, equations~\ref{eq:mean} and \ref{eq:mom2} can be used to model the probability distribution
\begin{equation}
    P(\tilde{\rho}|\rho)=\frac{1}{\sqrt{2\pi\text{Var}(\tilde{\rho}|\rho)}}\exp\left[-\frac{1}{2}\frac{(\tilde{\rho}-\langle\tilde{\rho}|\rho\rangle)^2}{\text{Var}(\tilde{\rho}|\rho)}\right].
\end{equation}
In practice, we are interested in the converse probability distribution, $P(\rho|\tilde \rho)$. Using Bayes theorem, we can evaluate the expectation value of $\rho$ given $\tilde \rho$, which is given by
\begin{equation}
    \langle\rho|\tilde{\rho}\rangle=\frac{\int\text{d}\rho\ P(\tilde{\rho}|\rho)P_\text{dust}(\rho)\rho}{\int\text{d}\rho\  P(\tilde{\rho}|\rho)P_\text{dust}(\rho)}.
    \label{eq:duster}
\end{equation}
We calculate this expectation value on a pixel by pixel basis to arrive at the final \duster\ map. Likewise, the second moment is given by
\begin{equation}
    \langle\rho^2|\tilde{\rho}\rangle=\frac{\int\text{d}\rho\ P(\tilde{\rho}|\rho)P_\text{dust}(\rho)\rho^2}{\int\text{d}\rho\  P(\tilde{\rho}|\rho)P_\text{dust}(\rho)},
\end{equation}
which we can use to determine the error in the \duster\ map on a pixel by pixel basis via
\begin{equation}
    \srho^2 \equiv {\rm Var}(\rho|\tilde \rho) = \avg{\rho^2|\tilde\rho} - \avg{\rho|\tilde\rho}^2.
\end{equation}
We emphasize that these are pixel-by-pixel marginalized uncertainties, and that in its present state, \duster\ is unable to generate correlated posteriors that can fully characterize the spatial structure of the statistical uncertainties.  We postpone the development of these capabilities to future work.

To follow standard conventions, we can now convert our final dust map $\rho$ into the extinction field $\calE$ by multiplying $\rho$ by $\overline{\calE}=28\ \mathrm{mmag}$, evaluated using some reference map (in our case, SFD) as averaged over the \duster\ footprint.  The map uncertainties are similarly scaled by this factor.

\section{Simulation Tests}
\label{sec:simulationtests}
\subsection{Statistical Performance}
\label{sec:statperf}
\begin{figure*}[tbp]
\centering 
\includegraphics[width=.8\textwidth]{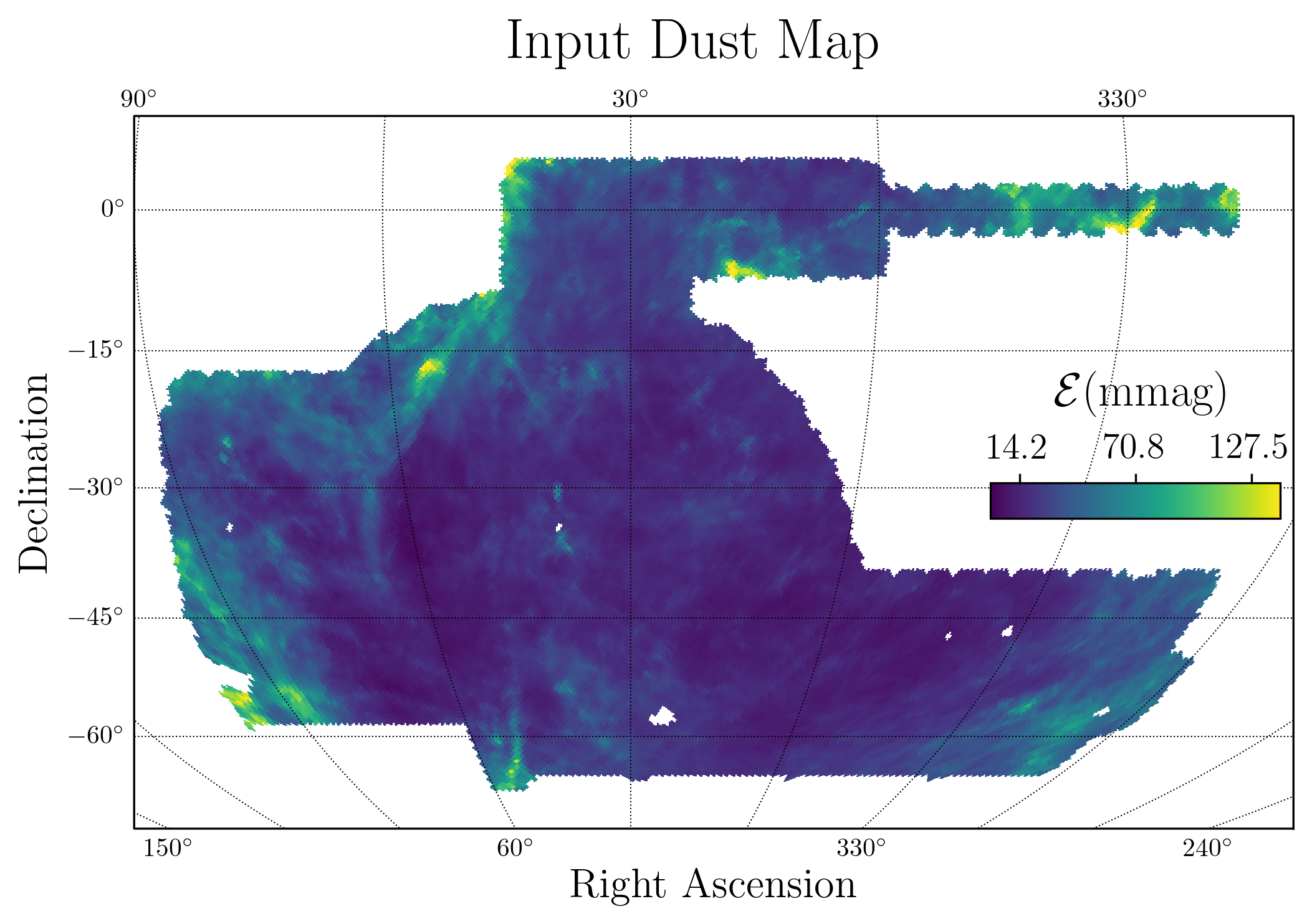}
\vspace{24pt}
\includegraphics[width=.8\textwidth]{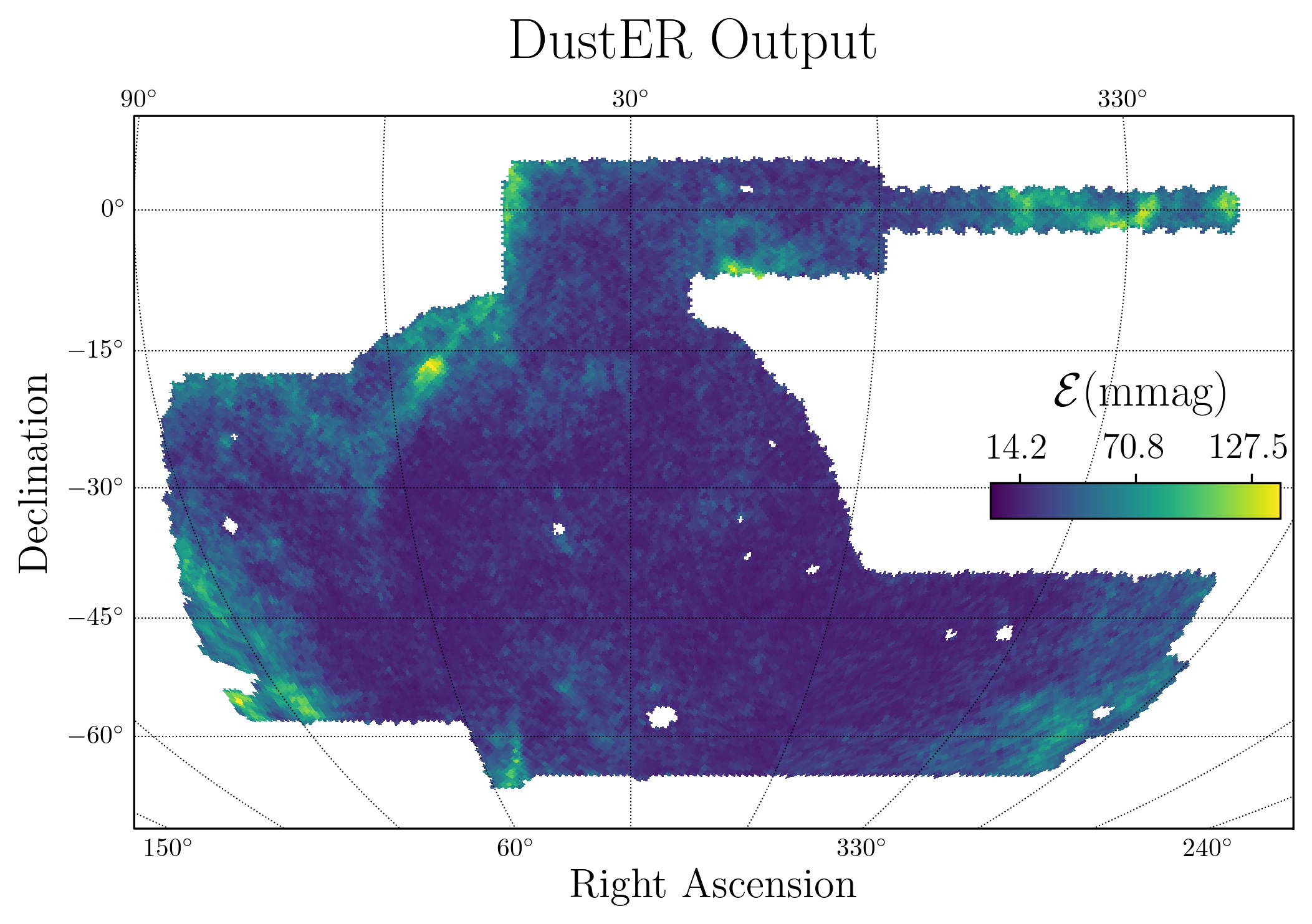}
\caption{Comparison between the input SFD dust map used to redden the simulated galaxies in \cardinal\ (top panel), to the dust map constructed using the \duster\ algorithm (bottom panel). Both maps are in units of mmag. The color scale of both maps is the same.}
\label{fig:maps}
\end{figure*}

\begin{figure}[tbp]
\centering 
\includegraphics[width=.45\textwidth]{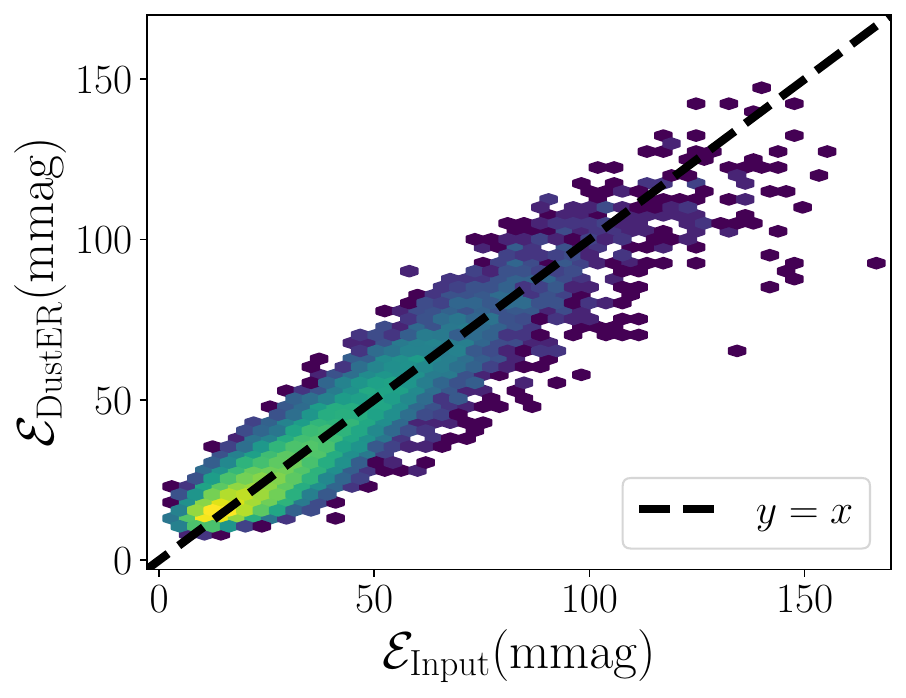}
\hfill
\includegraphics[width=.45\textwidth]{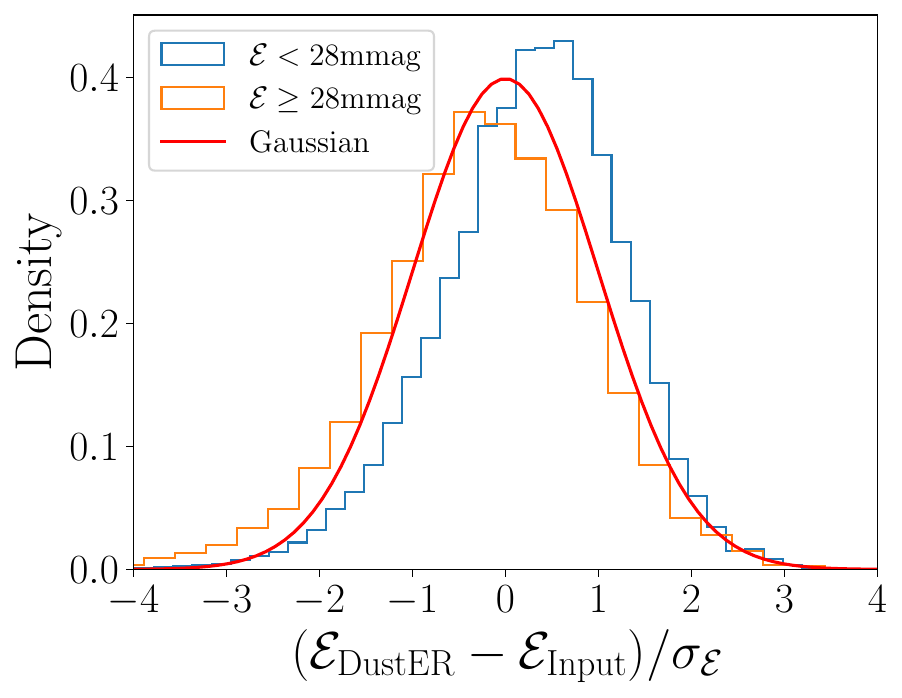}
\caption{{\bf Left:} \duster\ density as a function of the input (SFD) dust density in the \cardinal\ simulation.  The black dashed lines corresponds to the $y=x$ line.  {\bf Right:} A histogram of the difference between the \duster\ output and the input dust map (SFD), in units of the estimated uncertainty in \duster.  The two histograms are for pixels with $\mathcal{E}< 28\ \mathrm{mmag}$ (blue) and $\mathcal{E}\geq28\ \mathrm{mmag}$ (orange). The red solid curve is a Gaussian of zero mean and unit variance.}
\label{fig:map_performance}
\end{figure}

We apply the \duster\ algorithm described above to our simulated data set. Figure~\ref{fig:maps} compares the input SFD dust map used to redden all the galaxies in our simulation (top panel) with the \duster\ map obtained using equation~\ref{eq:duster} (bottom panel).  Visually, the two maps are in excellent agreement.  This agreement is better quantified in Figure~\ref{fig:map_performance}. The left panel plots the \duster\ density in the $y$-axis against the input (SFD) dust map in the $x$-axis.  We see that the \duster\ values scatter around the $y=x$ line (black dashed line), demonstrating that the \duster\ map is an unbiased estimate of the true map.

The scatter in the \duster\ map about the true map is shown in the right panel of Figure~\ref{fig:map_performance}.  A key feature of the \duster\ maps is that they come with statistical uncertainties.  In the panel, we plot histograms of the difference between the \duster\ map and the input SFD map in units of the reported error $\sigma_\mathcal{E}$. We show two histograms: one for $\mathcal{E} < 28\ \mathrm{mmag}$ (blue), and one for $\mathcal{E}\geq 28\ \mathrm{mmag}$ (orange). Also shown for reference is a Gaussian distribution of zero mean and unit variance.  The reason we show two histograms split by dust density is because the condition that $\mathcal{E}\geq 0$ forces the errors in the \duster\ map to be non-Gaussian in low extinction regions, leading to the skewed distribution of residuals for $\mathcal{E}< 28\ \mathrm{mmag}$ (i.e., $\rho<1$ which is below average).  In regions with a high dust density, the errors become approximately Gaussian.

\begin{figure}[tbp]
\centering
\begin{overpic}[width=\linewidth]{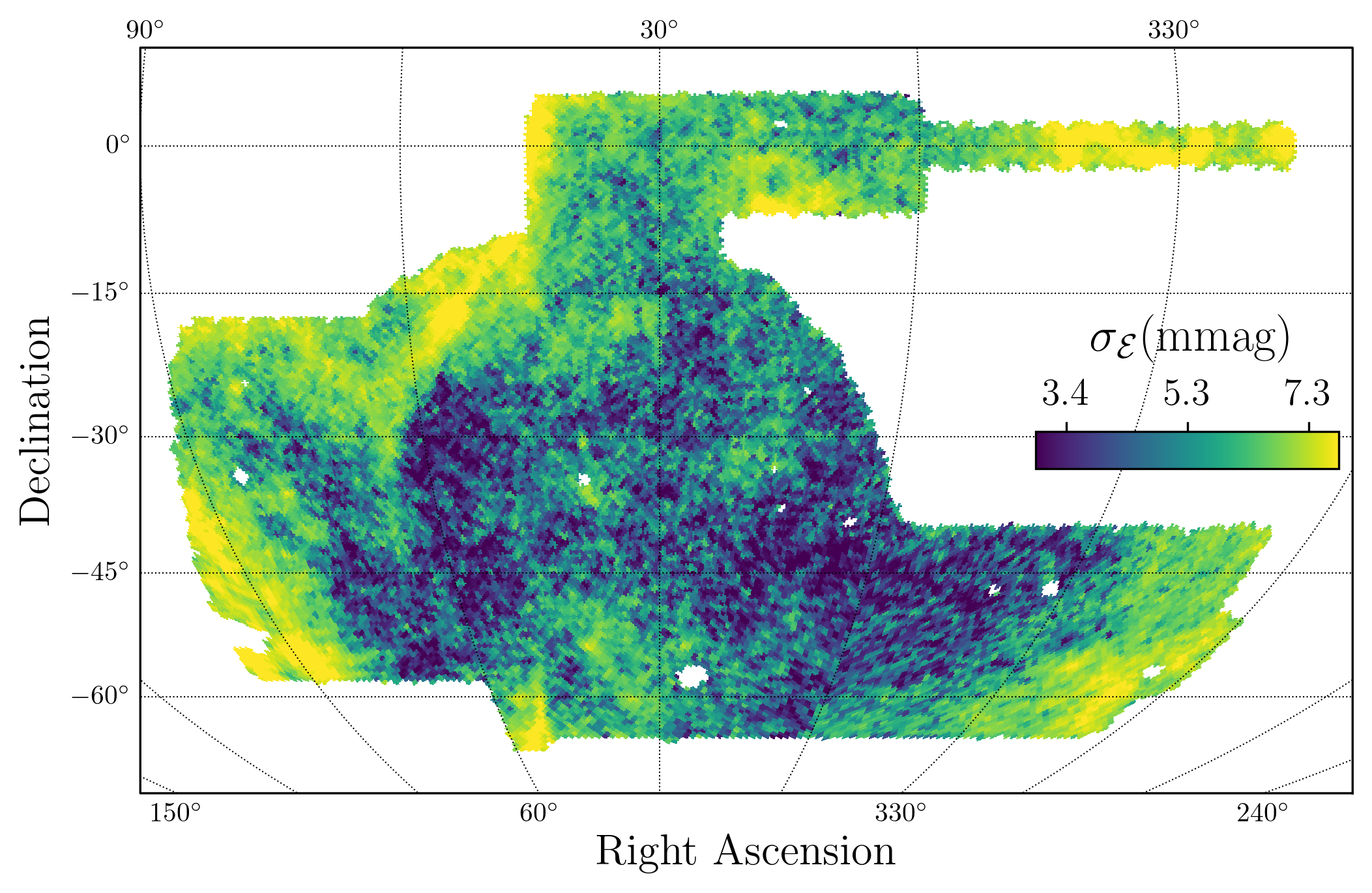}
    \put(0,45){%
        \fcolorbox{black}{white}{%
            \includegraphics[width=0.28\linewidth]{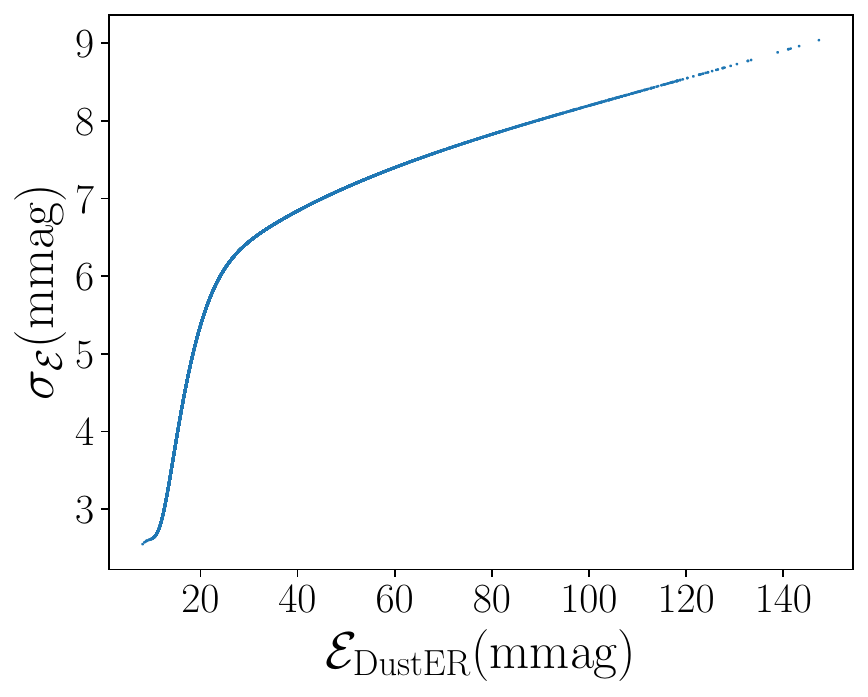}%
        }%
    }
\end{overpic}
\caption{The uncertainty in the recovered dust density $\sigma_\mathcal{E}$ as estimated using the \duster\ algorithm. The noise map is strongly correlated with the input dust map. The inset in the upper left shows the error in the recovered dust density as a function of the recovered dust density.}
\label{fig:errmap}
\end{figure}

Figure~\ref{fig:errmap} shows the estimated uncertainty $\sigma_\mathcal{E}$ as a function of position on the sky. We see the error map is highly correlated with the dust map itself.  This is better illustrated in the inset in the top-left of the plot, which shows the error $\sigma_\mathcal{E}$ in the \duster\ map as a function of the \duster\ dust density. Note the inset is actually a scatter plot: we do not impose a smooth relation between $\calE$ and its error.  The sharp drop in statistical uncertainty as one approaches $\mathcal{E}=0$ is due to the physical condition that $\mathcal{E}\geq 0$.

Table~\ref{tab:metrics_table} summarizes the global performance of \duster\ in our fiducial simulation. The \duster\ map recovers the input SFD map with a mean bias of $0.5\,\mathrm{mmag}$, an rms scatter of $6.3\,\mathrm{mmag}$, and a Pearson correlation  coefficient $r=0.94$.

\begin{figure}[tbp]
\centering 
\includegraphics[width=.65\textwidth]{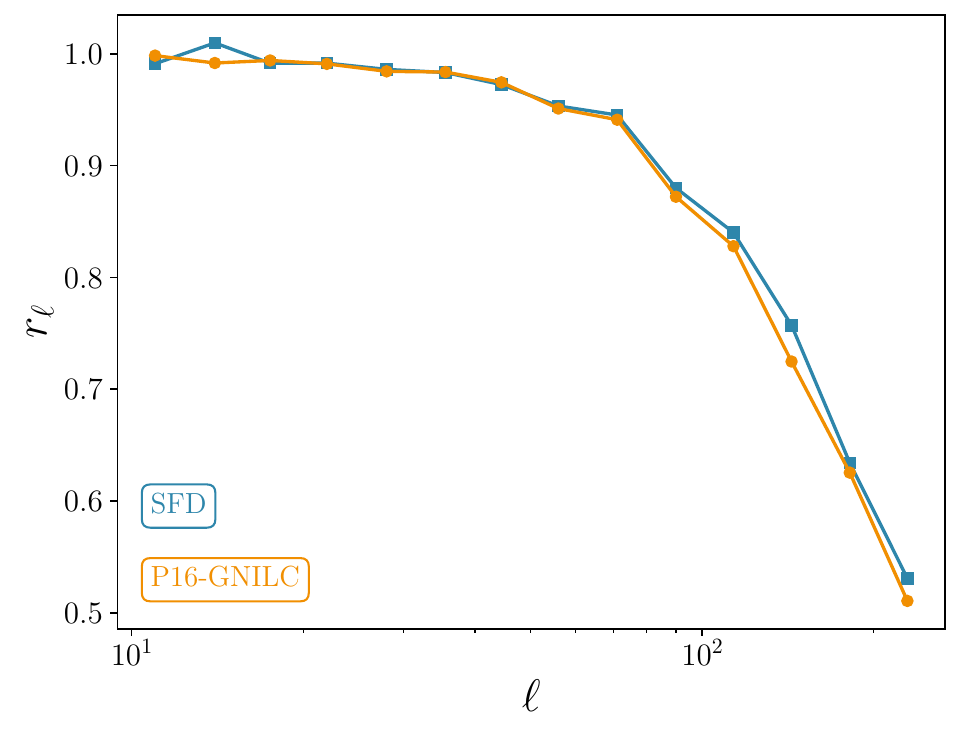}
\caption{Cross-correlation coefficient $r_\ell$ between the \duster\ map and the input dust map used to redden the simulated catalog, either SFD (blue) or P16-GNILC (orange). Both runs use the same one-point prior for the one point function of $\calE$. The recovered maps are nearly perfectly correlated with the input on large scales, and the correlation decreases on smaller scales, where noise dominates the reconstruction. The two curves are nearly indistinguishable, demonstrating that the mismatch between the SFD one-point function prior and the P16-GNILC one-point function has little impact on the quality of the reconstructed dust map.  Values slightly above unity at low $\ell$ reflect the noise of the binned pseudo-$C_\ell$ estimates.}
\label{fig:corr_coeff_input_output}
\end{figure}

To characterize the agreement as a function of angular scale, we compute
the cross-correlation coefficient
\begin{equation}
    r_\ell = \frac{C_\ell^{\rm DT}}{\sqrt{C_\ell^{\rm DD}\,C_\ell^{\rm TT}}},
    \label{eq:corr_coeff}
\end{equation}
where $C_\ell^{\rm DD}$ and $C_\ell^{\rm TT}$ are the auto-power spectra of the \duster\ map and the true (input) dust map, respectively, and $C_\ell^{\rm DT}$ is their cross-power spectrum. All spectra are pseudo-$C_\ell$ estimates computed with \textsc{NaMaster}\footnote{https://namaster.readthedocs.io} \citep{pseudocl}. Figure~\ref{fig:corr_coeff_input_output} shows that the \duster\ map is nearly perfectly correlated with the input map on large scales. The correlation decreases on smaller scales as the reconstruction becomes noise-dominated.

A natural concern is that \duster\ might simply reproduce its prior: the one-point function $P_{\rm dust}$ and the normalization $\overline{\mathcal{E}}$ are both derived from SFD, the same map used to redden the fiducial catalog. To test this, we apply \duster\ to the P16-GNILC-reddened catalog described in Section~\ref{sec:adding_dust}, keeping the SFD-based prior and normalization unchanged. The results are shown in the second and third rows of Table~\ref{tab:metrics_table}. If we compare the output \duster\ map to the true map in the simulation, namely the P16-GNILC map, we obtain a mean bias of $0.5\,\mathrm{mmag}$, an rms scatter of $7.1\,\mathrm{mmag}$, and $r=0.94$, comparable to the fiducial run. By contrast, the comparison of the \duster\ map to SFD is much noisier: rms of $8.3\,\mathrm{mmag}$, $r=0.89$. Evidently, even though \duster\ used an SFD-based prior, the \duster\ output is a much better tracer of the input P16-GNILC map. Critically, if we compare the correlation coefficient as a function of scale between the input map and \duster, the resulting correlation coefficient is nearly identical regardless of which dust map is used as the input to the simulations (see Figure~\ref{fig:corr_coeff_input_output}). These results provide strong evidence that the choice of prior used in \duster\ has little impact on the recovered dust map.

\begin{table}[tbp]
\centering
\begin{tabular}{|c|c|c|c|c|}
\hline
Input (reddening) map & Compared to & Mean bias (mmag) & RMS (mmag) & $r$ \\
\hline
SFD & SFD (truth) & $0.5$ & $6.3$ & $0.94$ \\
\hline
P16-GNILC & P16-GNILC (truth) & $0.5$ & $7.1$ & $0.94$ \\
\hline
P16-GNILC & SFD (prior only) & $0.5$ & $8.3$ & $0.89$ \\
\hline
\end{tabular}
\caption{\label{tab:metrics_table} Global performance of the \duster\ map
relative to a reference map over the simulation footprint: the mean bias $\langle\mathcal{E}_{\text{\duster}}-\mathcal{E}_{\rm ref}\rangle$, the rms difference, and the Pearson correlation coefficient $r$, computed over all $N_{\rm pix}\approx24{,}200$ valid pixels in each run. The input map is the map used to redden the simulated catalog, and the reference map is listed under ``Compared to''. In all runs, the one-point prior $P_{\rm dust}$ and the normalization $\overline{\mathcal{E}}$ are derived from SFD. The last row compares the P16-GNILC-reddened map to SFD, the source of the one-point function prior.}
\end{table}

\subsection{Systematics Tests}
\label{sec:systest}
The \duster\ map is obtained by using red-sequence galaxies as standard crayons.  It is therefore natural to consider whether the large-scale distribution of the Universe contaminates the \duster\ map.  We would like to address this question in a way that can be replicated with real data.  To that end, we consider the cross-correlation between \duster\ and a sample of red-sequence galaxies selected as described in Section~\ref{sec:redgals}.  However, unlike in Section~\ref{sec:redgals}, we will now select the galaxies after dereddening the raw photometry with the recovered \duster\ map.  We further bin the red galaxies in three redshift bins given by $z\in [0.20,0.40),[0.40,0.55),[0.55,0.70)$. We will use these same redshift bins throughout the rest of this paper, so we will henceforth refer to them as bins 1, 2, and 3, respectively. 

\begin{figure}[tbp]
\centering 
\includegraphics[width=.48\textwidth]{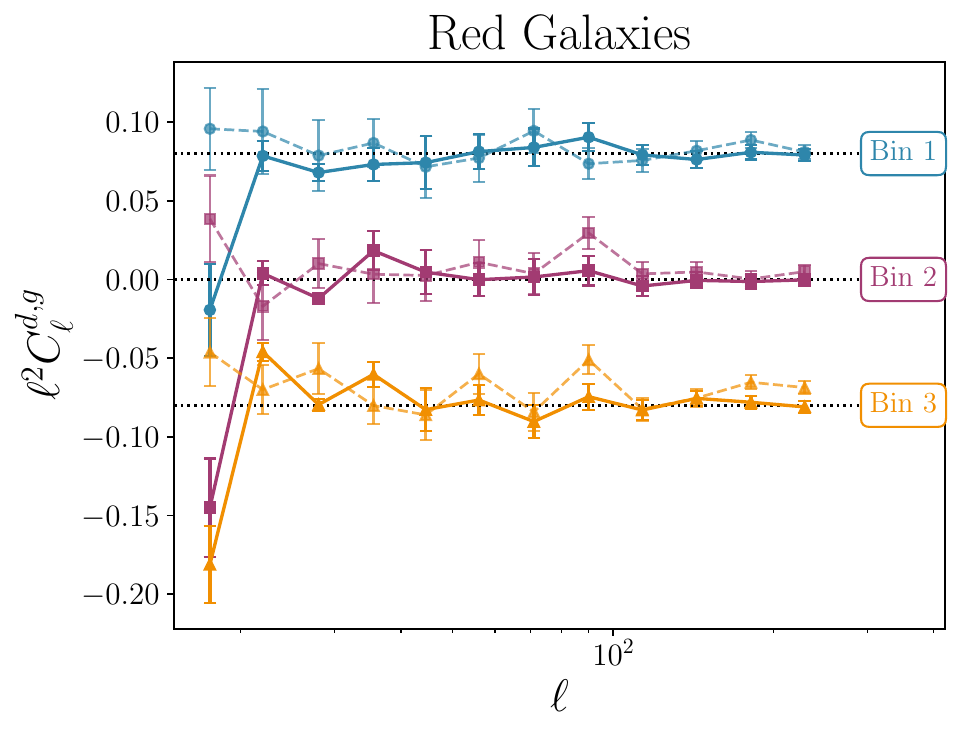}
\includegraphics[width=.48\textwidth]{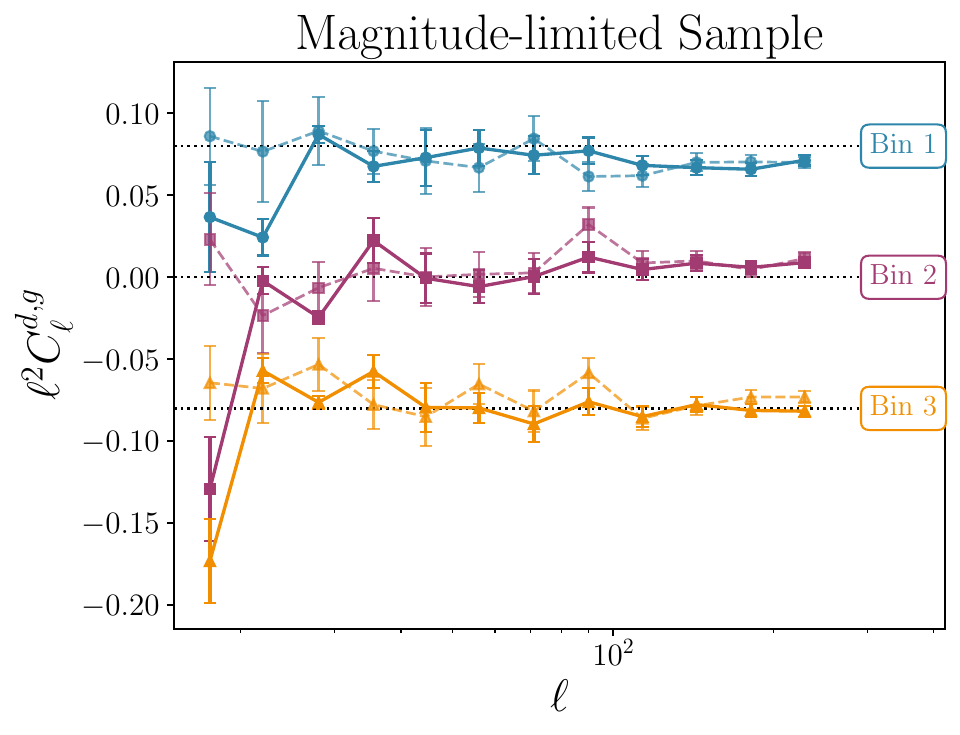}
\caption{{\bf Left:} The cross-power spectrum between the \duster\ map recovered from the simulations before (dashed line) and after (solid line) the LSS-subtraction using all three bins described in Section~\ref{sec:sysamel} and a sample of red-sequence galaxies selected as described in Section~\ref{sec:redgals}, with one exception: the galaxies are selected using magnitudes that were dereddened using the \duster\ map.  The galaxies are further binned by redshift into three sub-samples, as labeled.  Note the cross correlations of bins 1 and 3 have been shifted vertically to avoid overcrowding the plot. {\bf Right panel:} As the left panel, but for a magnitude limited sample of galaxies selected using the true magnitudes in the simulations (see Section~\ref{sec:systest}).}
\label{fig:sys1}
\end{figure}

The left panel in Figure~\ref{fig:sys1} shows the \duster--galaxy cross-correlation for each of the three redshift bins considered in this work. Statistical errors in the cross power spectrum are calculated following \cite{Krause16}, and the corresponding $\chi^2$ are reported in Table~\ref{tab:chisq}. Bin 3 deviates from zero ($\chi^2=37.75$, or $3.6\sigma$), while bins 1 and 2 are consistent with no cross-correlation. However, this alone is not sufficient evidence to unambiguously establish the presence of systematics in the \duster\ map, since the red-galaxies used for the test were selected using the \duster\ maps. 

To address this caveat, we cross-correlate the \duster\ map with a magnitude-limited sample comprising the brightest $5\%$ of \cardinal\ galaxies in the $i$ band
($m_i^{\rm true} < 21.95$), binned by spectroscopic redshift. The resulting cross correlations are shown in the right panel of Figure~\ref{fig:sys1}, and their $\chi^2$ are also reported in Table~\ref{tab:chisq}. The magnitude-limited sample shows excess correlation in bins 1 and 2 (2.7$\sigma$ and 2.3$\sigma$, respectively).

These results establish the existence of correlations between the \duster\ map and the galaxy density fields. However, the physical origin of these correlations remains ambiguous: they could arise either from LSS contamination in the \duster\ map itself, or from systematic biases in red galaxy selection due to imperfect dust corrections, or both. We address this question in the following section.

\begin{table}
\centering
\begin{tabular}{|c|c|c|}
\hline
$z$-bin & \duster\ $\times$ red galaxies & \duster\ $\times$ true-magnitude cut \\
\hline
$[0.20, 0.40)$ & 5.99 & 29.10 \\
$[0.40, 0.55)$ & 13.92 & 26.08 \\
$[0.55, 0.70)$ & 37.75 & 17.98 \\
\hline
\end{tabular}
\caption{$\chi^2$ values of the hypothesis that the dust--galaxy cross-power spectrum in Figure~\ref{fig:sys1} is zero.  The number of degrees of freedom is 12, so the 2$\sigma$ range for $\chi^2$ is [4.31,23.64].}
\label{tab:chisq}
\end{table}

\subsection{The Impact of Dust on Red-Galaxy Selection}
\label{sec:redgals_sel}
Before attempting to correct for the correlations detected in Section~\ref{sec:systest}, we must first establish their origin. Here we test whether red galaxy selection is systematically biased by dust or by survey depth variations. We consider seven different red galaxy samples, constructed as follows:
\begin{itemize}
    \item \textbf{Sample A}: We add photometric noise to \cardinal\ true magnitudes assuming a constant depth, and then select our usual red galaxy sample. By construction this sample contains no dust and no depth variation, so it is guaranteed to be uncorrelated with dust and serves as the control against which the remaining samples are compared.

    \item \textbf{Sample B}: We add dust attenuation to \cardinal\ true magnitudes based on the SFD map at a resolution of $\nside=4096$. We then add photometric noise assuming constant depth, and deredden the resulting magnitudes using the same SFD map at $\nside=4096$. Finally, we use the resulting catalog to select red galaxies. Sample B is identical to Sample A except that dust has been added and then removed, and therefore isolates whether dereddening magnitudes in the presence of photometric noise induces systematic density fluctuations in the galaxy density field.

    \item \textbf{Sample C}: Sample C is exactly like Sample B, except that magnitude dereddening is done by first averaging the SFD dust map to a resolution of $\nside=128$, and then using this low-resolution dust map to deredden galaxy magnitudes. The comparison between Samples B and C therefore isolates whether using a low resolution (but otherwise correct!) dust map can induce systematic density fluctuations in the galaxy density field.

    \item \textbf{Sample D}: We add photometric noise to \cardinal\ true magnitudes assuming variable depth, and then select our usual red galaxy sample. Sample D is identical to Sample A except for the depth treatment, and therefore isolates the effect of survey depth variations alone, in the complete absence of dust.

    \item \textbf{Sample E}: Sample E is exactly like Sample B, except that photometric noise is added assuming variable depth. Because Sample E differs from the control in both dust and depth, it measures the combined effect of the two rather than isolating either one.

    \item \textbf{Sample F}: Sample F is exactly like Sample C, except that photometric noise is added assuming variable depth; equivalently, it is Sample E with the dereddening map averaged to $\nside=128$.

    \item \textbf{Sample G}: Sample G is like Samples E and F, except that the low resolution map used to deredden galaxies is the \duster\ map itself. Unlike the SFD map used for the other samples, the \duster\ map carries its own reconstruction errors, so Sample G tests the practical consequence of dereddening a catalog with a \duster\ map.
\end{itemize}

We note that adding photometric noise to the simulations requires that we specify a depth map.  Here, we rely on a depth map constructed using raw (not de-reddened) photometry which we decorrelated from the SFD input dust map.  The reason is that the raw survey depth is a property of the survey, and is independent of the dust map.  However, we found that the DES Y3 depth maps allow for the dust map to contribute to the recovered depth. When one generates photometry using depth maps that are contaminated with dust, then the resulting red-galaxy selection is contaminated by dust (see Appendix~\ref{app:depth}). By using depth maps that are uncorrelated with dust, we are able to test whether \duster\ inherently leads to red-galaxy samples that are contaminated by dust.

To test for dust contamination of the galaxy density field, we cross-correlate each of the seven red galaxy samples with the input SFD dust map. The $\chi^2$ values of the zero cross-correlation hypothesis are summarized in Table~\ref{tab:chisq_sel}. All seven samples are consistent with zero cross-correlation at the $\approx 2\sigma$ level, demonstrating that at the DES survey depth, and for the range of galaxy magnitudes we consider, the spatial distribution of red galaxies is not impacted by dust, survey depth variations, or dereddening with the \duster\ map.  Appendix~\ref{app:depth} demonstrates that if the depth map meant to characterize the raw-depth photometry of the survey is incorrectly contaminated by dust, this does result in contamination of the large-scale structure probed by the red-sequence galaxies. 

\begin{table}
\centering
{\setlength{\tabcolsep}{4pt}
\small
\begin{tabular}{|c|c|c|c|c|c|c|c|}
\hline
$z$-bin & SFD $\times$ A & SFD $\times$ B & SFD $\times$ C & SFD $\times$ D & SFD $\times$ E & SFD $\times$ F & SFD $\times$ G \\
\hline
$[0.20, 0.40)$ &  6.82 & 10.32 &  9.29 & 5.82 & 10.89 & 9.97 & 6.10 \\
$[0.40, 0.55)$ &  9.07 &  9.34 &  8.73 & 8.50 & 9.14 & 8.71 & 11.75 \\
$[0.55, 0.70)$ & 18.37 & 21.23 & 21.48 & 18.05 & 22.77 & 22.66 & 23.69 \\
\hline
\end{tabular}}
\caption{$\chi^2$ values computed from the cross-power spectra between the input SFD dust map and each of the red galaxy samples. The number of degrees of freedom is 12, so the $2\sigma$ range for $\chi^2$ is [4.31,23.64]. All seven samples are consistent with zero at the $\approx 2\sigma$ level.}
\label{tab:chisq_sel}
\end{table} 

\begin{figure}[tbp]
\centering 
\includegraphics[width=.65\textwidth]{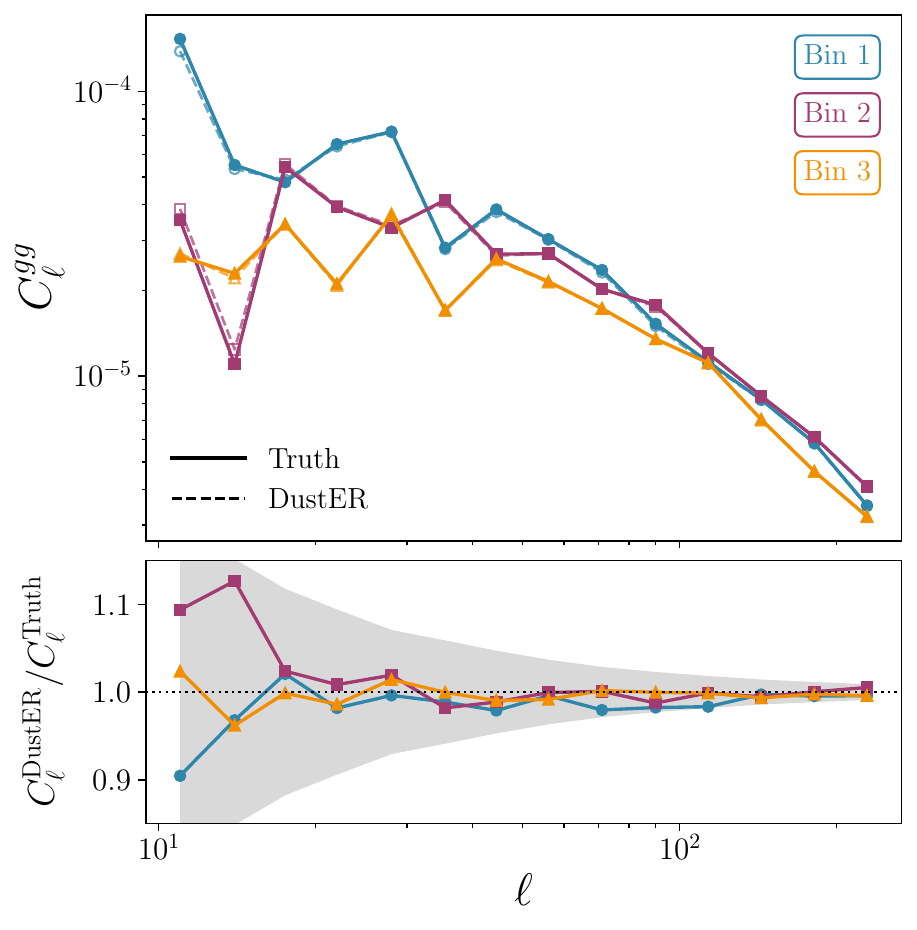}
\caption{{\bf Top:} Angular power spectra of red galaxies selected
after dereddening with the input SFD map at $\nside=4096$ (Sample E, solid)
and with the \duster\ map (Sample G, dashed), for each redshift bin.
{\bf Bottom:} Ratio of the two spectra. The gray band shows the
full-sky cosmic-variance uncertainty on $C_\ell^{gg}$. Because both samples trace
the same underlying density field, the uncertainty on the ratio itself is
smaller than this band; the band indicates the size of any bias relative to
the statistical error of a clustering measurement.}
\label{fig:clgg}
\end{figure}

We also test whether the LSS contamination of the \duster\ map propagates into the clustering of red galaxies selected after dereddening with it. Figure~\ref{fig:clgg} compares the angular power spectrum $C_\ell^{gg}$ of Sample G to that of Sample E, which is selected identically but dereddened with the input SFD map at $\nside = 4096$. For $\ell\gtrsim 20$ the two spectra agree within $\approx 2\%$ in all three redshift bins, and at all multipoles the ratio lies well within the full-sky cosmic-variance uncertainty of $C_\ell^{gg}$. The LSS contamination of the \duster\ map therefore does not measurably bias the clustering of the dereddened red-galaxy sample.

Having ruled out galaxy selection biases as the source of the correlations observed in Section~\ref{sec:systest}, we conclude that the \duster\ map itself is contaminated by LSS, consistent with the excess correlation found with the magnitude-limited sample. In the following section, we attempt to mitigate this contamination.

\subsection{Systematics Amelioration}
\label{sec:sysamel}

Having established the existence of LSS contamination in the \duster\ map, we test whether we can remove this contamination using an LSS-subtraction method based on the work by \citet{Chiang23}.  Let $\drhoD$ denote the density fluctuations of the ``dirty'' (i.e. LSS-contaminated) dust map.  We assume the LSS contamination takes the form of an additive systematic related to the density fluctuations of the galaxy sample used to create the \duster\ map.  This leads us to write the ``cleaned'' map $\drhoC$ as a linear combination of the dirty map and the galaxy density fields of interest.  Here, we consider the red-galaxy samples for each of the three redshift bins considered in the previous section.  In spherical harmonic space, the relation between these various maps is
\begin{equation}
    (\drhoC)_{\ell m} = (\drhoD)_{\ell m} + \sum_{i=1}^3 A^i_\ell (\delta_{g,i})_{\ell m}
    \label{eq:dclean}
\end{equation}
where $\delta_{g,i}$ is the galaxy density field for field $i$, and $A^i_\ell$ are constants which we will choose to remove the LSS signal from the \duster\ map.  

To find the appropriate values of the constant $A_\ell^i$, we consider the cross correlation between the cleaned \duster\ map and each of the three galaxy density fields. Multiplying equation~\ref{eq:dclean} by $(\delta_{g,j})^*_{\ell m}$, and averaging over $m$, and demanding the clean map exhibit no cross correlation with the galaxy density field $j$ (with $j\in\{1,2,3\}$)  we arrive at
\begin{equation}
    0 = C_\ell^{{\rm d}j} + \sum_{i=1}^3 A^i_\ell C_\ell^{ij}
\end{equation}
where $C_\ell^{{\rm d}j}$ is the cross-power spectrum between the dirty \duster\ map and galaxy bin $j$, and $C_\ell^{ij}$ is the cross-power spectrum between galaxy bin $i$ and galaxy bin $j$. The above system corresponds to three equations for each multipole $\ell$, which can be inverted to obtain the coefficients $A_\ell^i$. In practice, however, we cannot evaluate the cross-power spectra at every individual $\ell$ because of low signal-to-noise and the presence of masking. Instead, we compute pseudo-$C_\ell$ cross-power spectra using the \textsc{NaMaster} Python package  \citep{pseudocl}, which returns the cross-power at a set of effective multipoles $\ell_{\rm eff}$. We solve for $A_\ell^i$ at each $\ell_{\rm eff}$, and obtain $A_\ell^i$ at arbitrary 
$\ell$ by linearly interpolating the recovered values as a function of $\ell_{\rm eff}$.

\begin{table}
\centering
\begin{tabular}{|c|c|c|c|c|}
\hline
 & \multicolumn{2}{c|}{LSS Removal w/ Bins 1, 2, and 3} & \multicolumn{2}{c|}{LSS Removal w/ Bin 3 Only} \\
\hline
$z$-bin & Red galaxies $\chi^2$ & Magnitude cut $\chi^2$ & Red galaxies $\chi^2$ & Magnitude cut $\chi^2$ \\
\hline
$[0.20, 0.40)$ & 19.24 & 63.78 & 13.63 & 31.08 \\
$[0.40, 0.55)$ & 35.92 & 75.91 & 14.95 & 28.98 \\
$[0.55, 0.70)$ & 63.03 & 30.47 & 7.37 & 6.45 \\
\hline
\end{tabular}
\caption{\label{tab:sysremovetab} $\chi^2$ values of the hypothesis that the dust--galaxy cross-power spectrum is zero after large-scale structure removal using either all three redshift bins (left), or just redshift bin 3 (right, see Section~\ref{sec:sysamel} for details). The number of degrees of freedom is 12, so the $2\sigma$ range for $\chi^2$ is [4.31,23.64]. Using all three bins over-corrects the map, increasing the cross-correlation in every bin; removing bin 3 only reduces the cross-correlation in bin 3, but not in bins 1 and 2.}
\end{table}

The results in Table~\ref{tab:sysremovetab} and Figure~\ref{fig:sys1} show that the cleaning does not remove the contamination. When all three redshift bins are used, the $\chi^2$ of the cross-correlation increases in every bin, for both the red-galaxy and the magnitude-limited samples, indicating that the map has been over-corrected. Removing the contribution of bin 3 alone largely avoids this over-correction and substantially reduces the cross-correlation in bin 3, but the residual contamination in bins 1
and 2 is comparable to or higher than that in the uncleaned map. That is, the cleaning procedure is not effective at debiasing the \duster\ map.

\section{Summary and Discussion}
\label{sec:sumanddisc}
We have developed \duster, a photometric implementation of the standard crayon method to reconstruct Galactic dust reddening maps from red-sequence galaxies. The algorithm uses an empirically calibrated color--redshift relation to predict intrinsic galaxy colors as a function of redshift, and interprets coherent spatial color residuals as fluctuations in the dust extinction field. Because the red-sequence calibration and photometric redshifts are themselves affected by reddening, the initial reconstruction is both biased and noisy. \duster\ mitigates these effects through Wiener filtering and a statistical debiasing procedure that leverages the empirically determined one-point distribution of Galactic dust. The final dust map is obtained as the posterior expectation value of the extinction field, with pixel-level uncertainties derived from the posterior variance.

We validated \duster\ using simulated data based on the \cardinal\ catalog, in which dust attenuation is applied using the SFD dust map in our fiducial simulation. After debiasing, the recovered \duster\ map is statistically unbiased relative to the input dust field, with a mean bias of $0.5\,\mathrm{mmag}$ and an rms scatter of $6.3\,\mathrm{mmag}$. The reported uncertainties behave as expected: they are approximately Gaussian in the high-extinction regime and become non-Gaussian at low extinction due to the physical constraint $\calE\geq0$. The recovered map is nearly perfectly correlated with the input map on large scales, with the correlation decreasing on small scales where the reconstruction is noise-dominated. Because the one-point prior and normalization used by \duster\ are derived from SFD, we also reddened the simulation with the P16-GNILC dust map while keeping the SFD-based prior. The performance is nearly unchanged, and the recovered map agrees better with P16-GNILC than with SFD, demonstrating that \duster\ tracks the true dust field rather than reproducing its prior.

A key result of our analysis is the detection of statistically significant cross-correlations between the reconstructed \duster\ map and galaxy density fields in our simulations. To identify their origin, we performed a series of controlled tests on simulated red-galaxy samples with and without dust, including samples selected from catalogs that were reddened and then dereddened, either with the input dust map or with the \duster\ map itself. In none of these samples do we detect a correlation between the red-galaxy density field and the input dust map: at the depth of DES, red-galaxy selection from dereddened photometry is not measurably affected by dust, by survey depth variations, or by dereddening with the \duster\ map. Consequently, the non-zero \duster--red galaxy cross-correlation indicates that the \duster\ map itself is contaminated by large-scale structure. This is corroborated by a magnitude-limited sample selected on true magnitudes, which is independent of dust and depth by construction, yet correlates with \duster\ at $2.7\sigma$ and $2.3\sigma$ in the first two redshift bins, respectively. Nevertheless, this contamination does not propagate measurably into the angular clustering of red galaxies selected after dereddening with \duster: their angular power spectrum agrees with that obtained using the true dust map well within full-sky cosmic variance.

We explored a linear LSS subtraction scheme and found that it does not remove the contamination: using all three redshift bins over-corrects the map, while using only the highest-redshift bin reduces the cross-correlation in that bin but not in the others. Thus, the proposed linear cleaning algorithm fails. In future work, we plan to develop a forward-modeling, field-level inference framework in which the dust and galaxy density fields are modeled jointly within a hierarchical Bayesian approach, so that their correlations can be explicitly parameterized and marginalized over rather than removed by template subtraction. We anticipate this approach will yield posterior samples of the dust map and its uncertainties, with the maps being free of large-scale-structure correlations.

We plan to apply \duster\ to real photometric survey data, including observations from the Dark Energy Survey and the Dark Energy Camera All Data Everywhere (DECADE) \cite{Decade} program. These data will enable the construction of high-fidelity, purely photometric dust maps over large sky areas. The resulting reconstructions will provide an independent reddening-based perspective on Galactic extinction and will allow detailed comparisons with emission-based maps such as SFD and P16-GNILC. 

\section*{AI Disclosure}

The authors used Claude Opus 5.5 \citep{claude_opus55} to assist with editing the manuscript, checking internal consistency of the text, tables, and cross-references, and suggesting wording for sections. All scientific content, analysis, and results are the authors' own, and the authors reviewed and take full responsibility for the final text.

\acknowledgments

IEB, ER, and SSB were funded in part by NSF Grant 2306667. ER was also supported by DOE grant DE-SC0009913. ESR's work was performed in part under DOE Contract DE-AC02-76SF00515.

\appendix
\section{Dust Imprint in the Survey Depth Maps}
\label{app:depth}

We found the DES Y3 depth maps correlate with dust, even after correcting the depth maps so as to reflect the raw rather than dereddened photometry. Note this correlation must be spurious, and appears because of how the depth maps are built (see \citep{depth}). If we generate photometry using these maps in our simulations, we do find significant cross-correlation between red-sequence galaxies and the input dust map to the simulations, demonstrating LSS contamination by the dust map. 

This effect is clearly visible in the variable-depth samples of Section~\ref{sec:redgals_sel}. Table~\ref{tab:depth_before_after} compares the $\chi^2$ of the SFD--red galaxy cross-power spectra obtained with the original and corrected depth maps. In particular, Sample D, which contains no dust attenuation at all, was correlated with SFD at the $\approx 2.3\sigma$ level when the original depth maps were used. This is only possible if the depth maps themselves are correlated with the dust.

For these reasons, we perform the template subtraction method from Section~\ref{sec:sysamel} to remove dust contamination from the depth and effective-time-exposure maps.  It is these decorrelated maps that we used for our tests in the main body of the text.  With these corrected
depth maps, all variable-depth samples are consistent with zero at the $\approx 2\sigma$ level (Table~\ref{tab:depth_before_after}, bottom rows; see also Table~\ref{tab:chisq_sel}). 

\begin{table}[tbp]
\centering
\begin{tabular}{|c|c|c|c|c|}
\hline
$z$-bin & SFD $\times$ D & SFD $\times$ E & SFD $\times$ F & SFD $\times$ G \\
\hline
\multicolumn{5}{|c|}{Original depth maps (dereddened photometry)} \\
\hline
$[0.20, 0.40)$ & 17.09 & 30.94 & 28.98 & 16.20 \\
$[0.40, 0.55)$ & 11.37 & 17.55 & 18.21 & 28.60 \\
$[0.55, 0.70)$ & 25.71 & 31.04 & 29.74 & 34.26 \\
\hline
\multicolumn{5}{|c|}{Corrected depth maps (observed photometry)} \\
\hline
$[0.20, 0.40)$ & 5.82 & 10.89 & 9.97 & 6.10 \\
$[0.40, 0.55)$ & 8.50 & 9.14 & 8.71 & 11.75 \\
$[0.55, 0.70)$ & 18.05 & 22.77 & 22.66 & 23.69 \\
\hline
\end{tabular}
\caption{$\chi^2$ of the cross-power spectra between the input SFD map and the variable-depth red galaxy samples of Section~\ref{sec:redgals_sel}, using the original (dust-contaminated) and corrected depth maps. The number of degrees of freedom is 12, so the $2\sigma$ range for $\chi^2$ is [4.31, 23.64].}
\label{tab:depth_before_after}
\end{table}

\bibliography{sample}

\end{document}